\documentclass{LMCS}

\def\doi{8 (1:11) 2012}
\lmcsheading%
{\doi}
{1--32}
{}
{}
{Jun.~10, 2011}
{Feb.~27, 2012}
{}

\usepackage{enumerate,hyperref}
\usepackage{multicol}

\usepackage{amssymb}
\usepackage{prooftree}
\usepackage{soul}

\def\eg{{\em e.g.}}
\def\cf{{\em cf.}}
\def\ie{{\em i.e.}}

\newcommand{\tzero}{{\overline{0}}}

\newcommand{\val}[1]{[\![#1]\!]}
\newcommand{\ve}[1]{\mathrm{\textbf{#1}}}

\newcommand{\true}{\mathrm{true}}
\newcommand{\false}{\mathrm{false}}
\newcommand{\type}{\colon\!}
\newcommand{\Sc}{\mathcal{S}}

\newcommand{\B}{\mathcal{B}}
\newcommand{\V}{\mathcal{V}}

\newcommand{\lasf}{\ensuremath{\lambda 2^{la}}}
\newcommand{\rulesf}{^\triangleleft}
\newcommand{\mapsf}{^\natural}
\newcommand{\Tsf}{\ensuremath{\mathbb{T}(\lasf)}}
\newcommand{\nf}[1]{\ensuremath{\mathrm{nf}(#1)}}
\newcommand{\SAT}{\ensuremath{\mathrm{SAT}}}
\newcommand{\FV}{\ensuremath{\mathrm{FV}}}
\newcommand{\SN}{\ensuremath{\mathrm{SN}}}
\newcommand{\scalar}{\emph{Scalar}}

\begin{document}

\title{A System $F$ accounting for scalars}

\author[P.~Arrighi]{Pablo Arrighi\rsuper a}
\address{{\lsuper a}ENS-Lyon, Laboratoire LIP and Universit\'e de Grenoble, Laboratoire LIG, France}
\email{pablo.arrighi@imag.fr}

\author[A.~D\'iaz-Caro]{Alejandro D\'iaz-Caro\rsuper b}
\address{{\lsuper b}Universit\'e de Grenoble, Laboratoire LIG and Universit\'e Paris-Nord, Laboratoire LIPN, France}
\email{alejandro.diaz-caro@lipn.univ-paris13.fr}
\thanks{{\lsuper b}This work was partially supported by grants from DIGITEO and R\'egion \^Ile-de-France}

\titlecomment{An extended abstract of an earlier version of this paper has appeared in \cite{ArrighiDiazcaroQPL09}}

\begin{abstract}  
The algebraic lambda-calculus and the linear-algebraic lambda-calculus extend the lambda-calculus with the possibility of making arbitrary linear combinations of terms. In this paper we provide a fine-grained, System $F$-like type system for the linear-algebraic lambda-calculus. We show that this ``\scalar'' type system enjoys both the subject-reduction property and the strong-normalisation property, our main technical results. The latter yields a significant simplification of the linear-algebraic lambda-calculus itself, by removing the need for some restrictions in its reduction rules. But the more important, original feature of the \scalar\ type system is that it keeps track of `the amount of a type' that is present in each term.  As an example of its use, we show that it can serve as a guarantee that the normal form of a term is barycentric, i.e that its scalars are summing to one. 
\end{abstract}

\maketitle

\section{Introduction}
\subsection*{(Linear-)algebraic \texorpdfstring{$\lambda$}{lambda}-calculi} A number of recent works seek to endow  the $\lambda$-calculus with a vector space structure. This agenda has emerged simultaneously in two different contexts.
\begin{iteMize}{$\bullet$}
  \item The field of \emph{Linear Logic} considers a logic of resources where the propositions themselves stand for those resources -- and hence cannot be discarded nor copied. When seeking to find models of this logic, one obtains a particular family of vector spaces and differentiable functions over these. It is by trying to capture back these mathematical structures into a programming language that Ehrhard and Regnier have defined the {\em differential $\lambda$-calculus} \cite{EhrhardRegnierTCS03}, which has an intriguing differential operator as a built-in primitive and and an algebraic module of the $\lambda$-calculus terms over natural numbers. Lately Vaux \cite{VauxRTA07,VauxMSCS09} has focused his attention on a `differential $\lambda$-calculus without differential operator', extending the algebraic module to positive real numbers. He obtained a confluence result in this case, which stands even in the untyped setting. More recent works on this {\em algebraic $\lambda$-calculus} tend to consider arbitrary scalars \cite{EhrhardLICS10,TassonTLCA09}.
  \item The field of \emph{Quantum Computation} postulates that, as computers are physical systems, they may behave according to the quantum theory. It proves that, if this is the case, novel, more efficient algorithms are possible \cite{ShorSIAM97,GroverSTOC96} -- which have no classical counterpart. Whilst partly unexplained, it is nevertheless clear that the algorithmic speed-up arises by tapping into the parallelism granted to us `for free' by the {\em superposition principle}; which states that if $\ve{t}$ and $\ve{u}$ are possible states of a system, then so is the formal linear combination of them $\alpha.\ve{t}+\beta.\ve{u}$ (with $\alpha$ and $\beta$ some arbitrary complex numbers, up to a normalizing factor). The idea of a module of $\lambda$-terms over an arbitrary scalar field arises quite naturally in this context. This was the motivation behind the {\em linear-algebraic $\lambda$-calculus} by Dowek and one of the authors~\cite{ArrighiDowekRTA08}, who obtained a confluence result which holds for arbitrary scalars and again covers the untyped setting. 
\end{iteMize}
These two languages are rather similar: they both merge higher-order computation, be it terminating or not, in its simplest and most general form (namely the untyped $\lambda$-calculus) together with linear algebra in its simplest and most general form also (the axioms of vector spaces). In fact they can simulate each other \cite{DiazcaroPerdrixTassonValironHOR10}. Our starting point is the second one, also referred to as {\em Lineal} in this work: because its confluence proof allows arbitrary scalars and because one has to make a choice. Whether the models developed for the first language, and the type systems developed for the second language, carry through to one another via their reciprocal simulations, is a current topic of investigation.

\subsection*{Other motivations to study (linear-)algebraic \texorpdfstring{$\lambda$}{lambda}-calculi}  The two languages are also reminiscent of other works in the literature:
\begin{iteMize}{$\bullet$}
\item \emph{Algebraic and symbolic computation.} The functional style of programming is based on the $\lambda$-calculus together with a number of extensions, so as to make everyday programming more accessible. Hence since the birth of functional programming there have been several theoretical studies on extensions of the $\lambda$-calculus in order to account for basic algebra (see for instance Dougherty's algebraic extension \cite{DoughertyIC92} for normalising terms of the $\lambda$-calculus) and other basic programming constructs such as pattern-matching \cite{CirsteaKirchnerLiquoriFOSSACS01,ArbiserMiquelRiosJFP09}, together with sometimes non-trivial associated type theories \cite{PetitTLCA09}. Whilst this was not the original motivation behind (linear-)algebraic $\lambda$-calculi, they could still be viewed as just an extension of the $\lambda$-calculus in order to handle operations over vector spaces  and make programming more accessible upon them.
 The main difference in approach is that the $\lambda$-calculus is not seen here as a control structure which sits on top of the vector space data structure, controlling which operations to apply and when. Rather, the $\lambda$-calculus terms themselves can be summed and weighted, hence they actually are vectors, upon which they can also act.
  \item \emph{Parallel and probabilistic computation.} The above intertwinings of concepts are essential if seeking to represent parallel or probabilistic computation as it is the computation itself which must be endowed with a vector space structure. The ability to superpose $\lambda$-calculus terms in that sense takes us back to Boudol's parallel $\lambda$-calculus \cite{BoudolIC94}. It may also be viewed as being part of a series of works on probabilistic extensions of calculi, \eg~\cite{BournezHoyrupRTA03,HerescuPalamidessiFOSSACS00} and \cite{deLiguoroPipernoIC95,DipierroHankinWiklickyJLC05,DalLagoZorzi11} for $\lambda$-calculus more specifically.
\end{iteMize}
Hence (linear-)algebraic $\lambda$-calculi can be seen as a platform for various applications, ranging from algebraic computation, probabilistic computation, quantum computation and resource-aware computation.

\subsection*{Imposing the existence of a norm in (linear-)algebraic \texorpdfstring{$\lambda$}{lambda}-calculi}  We do not explore any of the above-mentioned fields of applications specifically in this paper. The theory of vector spaces has many fields of applications, but also many theoretical refinements that must be studied in their own right. We take the view that, in the same way, the theory of vector spaces plus $\lambda$-calculus has got theoretical refinements that we must study. Moreover, some of these theoretical refinements are often pre-requisite in order to address other applications, as is notoriously the case with the notion of norm. For instance if we want to be able to interpret a linear combination of terms $\sum \alpha_i.\ve{t}_i$ as a probability distribution, we need to make sure that it has norm one. The same is true if we want to interpret $\sum \alpha_i.\ve{t}_i$ as quantum superposition, but with a different norm\footnote{Whereas it is clear already that \emph{Lineal} is a quantum $\lambda$-calculus, in the sense that any quantum algorithm can be expressed in this language \cite{ArrighiDowekRTA08}, the converse, alas, is not true, in the sense that some programs in \emph{Lineal} express evolutions which are not valid quantum algorithms. This is precisely because \emph{Lineal} does not restrict its vectors to be normalized $\sum {|\alpha_i|}^2 = 1$ and its applications to be isometries relative to this norm.}. Yet the very definition of a norm is difficult in our context: deciding whether a term terminates is undecidable; but these terms can produce infinities (\eg~a non terminating term which reduces to itself plus 1), hence convergence of the norm is undecidable. Related to this precise topic, Vaux has studied simply typed algebraic $\lambda$-calculus, ensuring convergence of the norm \cite{VauxRTA07,VauxMSCS09}. Tasson has studied model-theoretic properties of the {\em barycentric} fragment of this simply-typed algebraic $\lambda$-calculus. Ehrhard has proven \cite{EhrhardLICS10} the convergence of a Taylor series expansion of algebraic $\lambda$-calculus terms, via a System $F$ type system \cite{GirardSLS71,ReynoldsPS74} .

Therefore, it can be said that standard type systems provide part of the solution: they ensure the convergence of (the existence of the norm of) a term. And indeed it is not so hard to define a simple extension of System $F$ that fits {\em Lineal} -- just by providing the needed rules to type additions, scalar products and the null vector in some trivial manner (see Definition \ref{def:lasf}). In this paper we do provide a full-blown proof of strong normalisation for this type system. A byproduct of this result is that we are able to lift several conditions that were limiting the reduction rules of {\em Lineal} and still prove its confluence. This is because the purpose of these conditions was really to prevent indefinite forms from reducing (\eg\;$\ve{t}-\ve{t}$, with $\ve{t}$ a divergent term). Hence the result together with its byproduct make \emph{Lineal} into a simpler language.

\subsection*{Imposing a given norm in (linear-)algebraic \texorpdfstring{$\lambda$}{lambda}-calculi} Standard type systems only provide a part of the solution; they are unable for instance to impose that well-typed terms have normal form $\sum\alpha_i.\ve{t}_i$ with the constraint that $\sum \alpha_i=1$, a property which is referred to as barycentricity in the previous literature \cite{TassonTLCA09}. Our main contribution in this paper is to have managed a type system of a novel kind, \scalar, which has the ability to keep track of `the amount of a type' by summing the amplitudes of its contributing terms and reflecting this amount through scalars within the types. As an example application of the \scalar\ type system, we have demonstrated that the type system can be used to enforce the barycentric property (see section~\ref{sec:disc}), thereby specializing {\em Lineal} into a barycentric, higher-order $\lambda$-calculus. 

Endowing \emph{Lineal} with a non-trivial, more informative, fine-grained type system was quite a challenge, as the reader will judge. We believe that fine-grained type theories for these non-deterministic,  parallel, probabilistic or quantum extensions of calculi, which capture how many processes are in what class of states (with what weight), constitute truly promising routes towards relevant, less ad-hoc quantitative logics.

\subsection*{Plan} Section~\ref{sec:language} presents an overview of the linear-algebraic $\lambda$-calculus (\emph{Lineal}) \cite{ArrighiDowekRTA08}. Section~\ref{sec:scalar} presents the \scalar\ type system with its grammar, equivalences and inference rules. Section~\ref{sec:subjectreduction} shows the subject reduction property giving consistency to the system. Section~\ref{sec:strongnormalisation} shows the strong normalisation property for this system, allowing us to lift the above discussed restrictions in the reduction rules. In section~\ref{sec:disc} we formalize the type system $\mathcal{B}$ for barycentric calculi. Section~\ref{sec:conclusion} concludes. 

\section{Linear-algebraic \texorpdfstring{$\lambda$}{lambda}-calculus}\label{sec:language}
\subsection*{Intuitions} As a language of terms, {\em Lineal} is just the $\lambda$-calculus together with the possibility to make arbitrary linear combinations of terms: if $\ve t$ and $\ve r$ are two terms, $\alpha.\ve{t}+\beta.\ve{r}$ is also a term. Regarding the operational semantics, {\em Lineal} merges higher-order computation, be it terminating or not, in its simplest and most general form (the $\beta$-reduction of the untyped $\lambda$-calculus) together with linear algebra in its simplest and most general form also (the oriented axioms of vector spaces). Care must be taken, however, when merging these two families of reduction rules.

For instance the term $(\lambda x\,xx)(\alpha.\ve{t}+\beta.\ve{u})$ maybe thought of as reducing to $(\alpha.\ve{t}+\beta.\ve{u})(\alpha.\ve{t}+\beta.\ve{u})$ in a call-by-name-oriented view or to $\alpha.(\ve{t}\ve{t})+\beta.(\ve{u}\ve{u})$ in a call-by-value-oriented view, also compatible with the view that application should be bilinear, just as in linear algebra (\cf~\textit{Application rules}, below). Leaving both options open would break confluence, the second option was chosen, which entails restricting the $\beta$-reduction to terms not containing sums or scalars in head position (\cf~\textit{Beta reduction} rule, below).

Instead of introducing vector spaces via an oriented version of their axioms (\eg~$\alpha.\ve{u}+\beta.\ve{u}\to(\alpha+\beta).\ve{u}$), one could have decided to perform the $\beta$-reduction `modulo equality in the theory of vector spaces' (\emph{e.g.\ }$\alpha.\ve{u}+\beta.\ve{u}=(\alpha+\beta).\ve{u}$). But there is a good reason not to do that. Define a fixed point operator 
$$\ve{Y}=\lambda y\,(\lambda x\,(y+xx))(\lambda x\,(y+xx))$$
and consider a term $\ve{b}$ such that $\ve{Y}\ve{b}$ is $\beta$-equivalent to $\ve{b}+\ve{Y}\ve{b}$ and so on. Modulo equality over vector spaces, the theory would be inconsistent, as the term $\ve{Y}\ve{b}-\ve{Y}\ve{b}$ would then be equal to $\ve{0}$, but would also reduce to $\ve{b}+\ve{Y}\ve{b}-\ve{Y}\ve{b}$ and hence also be equal to $\ve{b}$. Instead, this problem can be fixed by restricting rules such as $\alpha.\ve{u}+\beta.\ve{u}\to(\alpha+\beta).\ve{u}$ to terms that cannot reduce forever (\cf~\textit{Factorisation rules}, below), matching the old intuition that indefinite forms `$\infty-\infty$' must be left alone. Moreover, oriented axioms of vector spaces define vector spaces and no more than vector spaces, just as well as the original axioms do, as was shown in \cite{ArrighiDowekRTA08}. Plus the orientation serves a purpose: it presents the vector in its canonical form. 

\subsection*{Definitions} The untyped {\em Lineal} calculus, as defined in \cite{ArrighiDowekRTA08}, is presented in Figure~\ref{fig:lineal}. 
Terms contain a subclass of \emph{base terms}, that are the only ones that can be substituted for a variable in a $\beta$-reduction step.
We use the convention that terms equal up to associativity and commutativity axioms on $+$ (that is an {\em AC-rewrite system}). In particular, we have $\ve t+\ve u=\ve u+\ve t$ and $\ve t+(\ve u+\ve r)=(\ve t+\ve u)+\ve r$. That rewriting modulo equational theory has been extensively studied~\cite{JouannaudKirchnerSIAM86,PetersonStickelACM81} to a high-level of rigour, and largely discussed in the context of this calculus in~\cite{ArrighiDowekRTA08}.
Scalars (denoted by $\alpha, \beta, \gamma,\dots$) are members of a commutative ring $(\Sc, +, \times, 0)$. Scalars can be defined as forming a rewrite system by itself \cite{ArrighiDowekRTA08}, however for the sake of simplicity, it is enough to consider them as members of a ring. For example, the term $6.\ve t$ and the term $(2\times 3).\ve t$ are considered to be the same term.
Also, $\alpha$-equivalence is considered implicitly, that is we rename variables when needed to avoid variable capturing. To be fully formal we would need to introduce these three equivalences, however they are standard so we do not mention them explicitly as long as it does not lead to confusion.
The parentheses associations are the standard. We add the following conventions:  $\alpha.\beta.\ve t$ stands for $\alpha.(\beta.\ve t)$ and $\ve t_1\dots\ve t_n+\ve r_1\dots\ve r_n$ stands for $(\ve t_1\dots\ve t_n)+(\ve r_1\dots\ve r_n)$. 

The confluence of this calculus has been formally proven in \cite{ArrighiDowekRTA08}.

The set of free variables of a term (notation:~$\FV({\ve t})$) is defined as expected. The operation of substitution on terms (notation:~$\ve{t}[\ve{b}/x]$) is defined as usual (\ie~taking care of renaming bound variables when needed in order to prevent variable capture), with $(\alpha.\ve t+\beta.\ve u)[\ve{b}/x]=\alpha.(\ve t[\ve{b}/x])+\beta.(\ve u[\ve{b}/x])$.  Finally, $\nf{\ve t}$ denotes the unique normal form of the term $\ve t$ and it is used only when it is clear that $\ve t$ has a unique normal form.
As usual, $\to^*$ is the application of zero or more reduction rules.
\begin{figure}[ht]
  \hrule\vspace{1pt}\hrule
  \begin{displaymath}
    \begin{array}[t]{l@{\hspace{1.5cm}}r@{\ ::=\quad}l}
      \text{\itshape Terms:} & \ve{r},\ve{t},\ve{u} & 
      \ve{b}\ |\ \ve{t}\ve{r}\ |\ \ve{0}\ |\ \alpha.\ve{t}\ |\ \ve{t}+\ve{r}\\
      \text{\itshape Base terms:} & \ve{b} & x\ |\ \lambda x\,\ve{t}
    \end{array}
  \end{displaymath}
\begin{tabular}{p{3.7cm}p{4.5cm}p{6cm}}
\emph{Elementary rules:}

$\ve{t}+\ve{0}\to\ve{t}$,

$0.\ve{t}\to\ve{0}$,

$1.\ve{t}\to\ve{t}$,

$\alpha.\ve{0}\to\ve{0}$,

$\alpha.(\beta.\ve{t})\to(\alpha\times\beta).\ve{t}$,

$\alpha.(\ve{t}+\ve{r})\to\alpha.\ve{t}+\alpha.\ve{r}$.
&
\emph{Factorisation rules:}

$\alpha.\ve{t}+\beta.\ve{t}\to(\alpha+\beta).\ve{t}$  (*),

$\alpha.\ve{t}+\ve{t}\to(\alpha+1).\ve{t}$  (*),

$\ve{t}+\ve{t}\to(1+1).\ve{t}$  (*).
\bigskip

\emph{Beta reduction:}

$(\lambda x\,\ve{t})\ve{b}\to\ve{t}[\ve{b}/x]$ (***).
&
\emph{Application rules:}

$(\ve{t}+\ve{r})\ve{u}\to\ve{t}\ve{u}+\ve{r}\ve{u}$ (**),

$\ve{u}(\ve{t}+\ve{r})\to\ve{u}\ve{t}+\ve{u}\ve{r}$ (**),

$(\alpha.\ve{t})\ve{r}\to\alpha.\ve{t}\ve{r}$ (*),

$\ve{r}(\alpha.\ve{t})\to\alpha.\ve{r}\ve{t}$ (*),

$\ve{0}\ve{t}\to \ve{0}$,

$\ve{t}\ve{0}\to \ve{0}$.
\end{tabular}
\medskip

\emph{Contextual rules:} If $\ve t\to\ve r$, then for any term $\ve{u}$ and variable $x$,
\begin{multicols}{3}
$\ve t\ve{u}\to\ve r\ve{u}$,\\
$\ve{u}\ve t\to\ve{u}\ve r$,\\
$\ve t+\ve{u}\to\ve r+\ve{u}$,\\
$\ve{u}+\ve t\to\ve{u}+\ve r$,\\
$\alpha.\ve t\to\alpha.\ve r$ and\\
$\lambda x\,\ve t\to\lambda x\,\ve r$.
\end{multicols}
\medskip

\begin{flushleft}
where $+$ is an associative-commutative ({\em AC}) symbol and\\
(*) these rules apply only if $\ve{t}$ is a closed normal term.\\
(**) these rules apply only if $\ve{t}+\ve{r}$ is a closed normal term.\\
(***) the rule applies only when $\ve{b}$ is a base term.

Restriction (***) is the one that limits the $\beta$-reduction, whereas restrictions (*) and (**) are those that avoid confluence problems related to infinities and indefinite forms, as discussed above. 
\end{flushleft}

  \hrule\vspace{1pt}\hrule
  \caption{Syntax and reduction rules of {\em Lineal}.}
  \label{fig:lineal}
\end{figure}

\begin{exa} One may be tempted to think $\lambda x\,2.x\sim 2.\lambda x\,x$. Consider the following reduction.
$(\lambda x\,2.x+\lambda x\,x)(3.\ve b_1+2.\ve b_2)$
 \begin{eqnarray*}
  &\to^*& 3.((\lambda x\,2.x)\ve b_1)+3.((\lambda x\,x)\ve b_1)+2.((\lambda x\,2.x)\ve b_2)+2.((\lambda x\,x)\ve b_2)\\
  &\to^*& 3.2.\ve b_1+3.\ve b_1+2.2.\ve b_2+2.\ve b_2\\
  &\to^*&6.\ve b_1+3.\ve b_1+4.\ve b_2+2.\ve b_2\\
  &\to^*&9.\ve b_1+6.\ve b_2
 \end{eqnarray*}
However while these terms behave the same in this case, they are different: the first one is the function (a base term) duplicating its argument, while the second one is twice the identity function. The display below is clarifying:
\begin{eqnarray*}
 (\lambda x\,y)(\lambda x\,2.x) &\to& y,\\
 (\lambda x\,y)(2.\lambda x\,x) &\to& 2.(\lambda x\,y)(\lambda x\,x)\to 2.y.
\end{eqnarray*}
\end{exa}

\section{The \scalar\ Type System}\label{sec:scalar}
The grammar of types, \cf~Figure~\ref{fig:types}, defines the set of types (denoted by $T,R,S$) and its syntactic subclass (denoted by $U,V,W$) of what we call \emph{unit types}. Note that the grammar for unit types does not allow for scalars except to the right of an arrow. More generally, note the novelty of having scalars at the level of types. We will see that these scalars keep track of the sum of the coefficients of the terms contributing to the type.

Type variables are denoted by~$X, Y$, etc.~and can only ever be substituted by a unit type. Contexts are denoted by $\Gamma,\Delta$, etc.~and are defined as sets $\{x\type U, \ldots\}$, where $x$ is a term variable appearing only once in the set and $U$ is a unit type. We usually omit the brackets of the set. The substitution of~$V$ for~$X$ in~$U$ is defined as usual and is written~$U[V/X]$. We sometimes use the vectorial notation~$U[\vec{V}/\vec{X}]$ for~$U[V_1/X_1]\cdots[V_n/X_n]$ if~$\vec{X}=X_1,\dots,X_n$ and~$\vec{V}=V_1,\dots,V_n$. In order to avoid capture, we consider that $X_i$ cannot appear free in $V_j$, with $j<i$. Also we assume that free and bound variables of a type are distinct. We also may abuse notation and say $\vec X\notin S$ meaning that none of the $X_i$ from $\vec X$ are in the set $S$. We write $\Gamma[U/X]$ to the substitution of $U$ for $X$ in each type of $\Gamma$. Also we write $(\alpha.T)[U/X]$ for $\alpha.T[U/X]$. $\FV(T)$ designates the set of free variables of the type $T$, defined in the usual manner,  and $\FV(\Gamma)$ is the union of the sets of free variables of each type in $\Gamma$. Scalars are denoted by $\alpha,\beta,\gamma\dots$ and are members of the same commutative ring $(\Sc, +,\times)$ as those of terms.

We also define an equivalence relation upon types as follows:

\begin{defi}\label{def:equiv}
For any $\alpha, \beta$ and $T$. We define the type equivalence $\equiv$ to be the least congruence such that
$$\bullet\ \alpha.\tzero\equiv\tzero\quad\bullet\ 0.T\equiv\tzero\quad\bullet\ 1.T\equiv T\quad\bullet\ \alpha.(\beta.T)\equiv(\alpha\times\beta).T\quad\bullet\ \forall X.\alpha.T\equiv\alpha.\forall X.T$$
\end{defi}

\begin{figure}[ht]
  \hrule\vspace{1pt}\hrule
 \begin{displaymath}
    \begin{array}[t]{l@{\hspace{1.5cm}}r@{\ ::=\quad}l}
      \text{\itshape Types:} & R,S,T & U\ |\ \forall X.T\ |\ \alpha.T\ |\ \tzero\\
      \text{\itshape Unit types:} & U,V,W & X\ |\ U\to T\ |\ \forall X.U\\
    \end{array}
  \end{displaymath}
  \medskip
  
  \begin{tabular}{c@{\hspace{0.25cm}}c@{\hspace{0.25cm}}c}
  \prooftree
  \justifies\Gamma, x\type{U}\vdash x\type{U}
  \using ax
  \endprooftree
  &
  \prooftree\Gamma\vdash\ve{t}\type T\qquad T\equiv S
  \justifies\Gamma\vdash\ve{t}\type S
  \using\equiv
  \endprooftree
  &
  \prooftree\Gamma \vdash\ve{t}\type\alpha.({U}\to {T}) \qquad \Gamma\vdash\ve{r}\type\beta.{U}
  \justifies\Gamma \vdash\ve{t}\ve{r}\type(\alpha\times\beta).{T}
  \using\to_E
  \endprooftree \\
  & & \\
  \prooftree\Gamma, x\type{U} \vdash\ve{t}\type{T}
  \justifies\Gamma \vdash \lambda x\,\ve{t}\type{U}\to {T}
  \using\to_I
  \endprooftree
  &
  \prooftree\Gamma\vdash\ve{t}\type\forall X.{T}
  \justifies\Gamma\vdash\ve{t}\type{T}[U/X]
  \using\forall_E
  \endprooftree
  &
  \prooftree\Gamma \vdash\ve{t}\type{T}\qquad X\notin \FV(\Gamma)
  \justifies\Gamma \vdash\ve{t}\type\forall X.{T}
  \using\forall_I
  \endprooftree \\
  & & \\
  \prooftree
  \justifies\Gamma\vdash\ve{0}\type\tzero
  \using ax_\tzero
  \endprooftree
  &
  \prooftree \Gamma\vdash\ve{t}\type\alpha.T \qquad \Gamma\vdash\ve{r}\type\beta.T
  \justifies \Gamma\vdash\ve{t}+\ve{r}\type(\alpha+\beta).T
  \using +_I
  \endprooftree
  &
  \prooftree \Gamma\vdash\ve{t}\type T
  \justifies\Gamma\vdash\alpha.\ve{t}\type\alpha.T
  \using s_I
  \endprooftree
  \end{tabular}
  \medskip
  
  \hrule\vspace{1pt}\hrule
  \caption{Types and typing rules of \scalar}
  \label{fig:types}
\end{figure}

The complete set of typing rules is shown in Figure~\ref{fig:types}. Splitting the grammar into general types and unit types is a necessary consequence of the fact that we want scalars in the types to reflect scalars in the terms (\eg~$\alpha.\lambda x\,\ve{t}$ should have a type $\alpha.U$). Indeed if we did not have the restriction on the left hand side of an arrow being a unit type, \ie~$U\to T$, then we would have types like $(\alpha.X)\to X$, which \emph{a priori} do not make sense, because abstractions receive only base terms as arguments. This could be fixed by adding the equivalence $(\alpha.A)\to B\equiv\alpha.(A\to B)$, making sure that $\alpha$ is non-zero. But still we would need to keep the $\to_E$ rule restricted to having a unit type on the left of the arrow, otherwise we would break the required correspondence between scalars-in-types and scalars-in-terms, \eg:
$$\prooftree\vdash\alpha.\lambda x\,x\type (\alpha.T)\to T\qquad\vdash\ve{t}\type\alpha.T
\justifies\vdash(\alpha.\lambda x\,x)\ve{t}\type T
\endprooftree
\qquad
\textrm{but }(\alpha.\lambda x\,x)\ve{t}\to^*\alpha.\ve{t} \textrm{ which should be of type  }\alpha^2.T$$

We want the scalars in the types to represent those in the terms, hence the rule $s_I$. The rule $+_I$ takes care of sums of terms and the term $\ve 0$ gets the special type $\tzero$ by an axiom.

Finally, let us go back to the application. The standard rule $\to_E$ from System $F$ needs to be consistent with the extra rules for application that we have on top of $\beta$-reduction in {\em Lineal}; namely the \textit{Application rules}:
\begin{multicols}{2}
\begin{enumerate}[(1)]
  \item\label{rule:linizq} $(\ve{t}+\ve{r})\ve{u}\to \ve{t}\ve{u}+\ve{r}\ve{u}$
  \item\label{rule:linder} $\ve{u}(\ve{t}+\ve{r})\to\ve{u}\ve{t}+\ve{u}\ve{r}$
  \item\label{rule:scalizq} $(\alpha.\ve{t})\ve{r}\to\alpha.\ve{t}\ve{r}$
  \item\label{rule:scalder} $\ve{r}(\alpha.\ve{t})\to\alpha.\ve{r}\ve{t}$
  \item\label{rule:0izq} $\ve{0}\ve{t}\to \ve{0}$
  \item\label{rule:0der} $\ve{t}\ve{0}\to \ve{0}$
\end{enumerate}
\end{multicols}
\noindent Note that the terms $\ve{t}$ and $\ve{r}$ in rules (\ref{rule:linizq}) and (\ref{rule:linder}) must now have the same type (up to a scalar) according to the rule $+_I$, so the type of $\ve{t}+\ve{r}$ is analogous to the type of $\alpha.\ve{t}$ in rules (\ref{rule:scalizq}) and (\ref{rule:scalder}). Also, the type for $\ve{0}$ in rules (\ref{rule:0izq}) and (\ref{rule:0der}) is the same as that of $0.\ve{t}$ if we take $\alpha=0$ in rules (\ref{rule:scalizq}) and (\ref{rule:scalder}). Thus we can focus our discussion on rules (\ref{rule:scalizq}) and (\ref{rule:scalder}).

By the rule (\ref{rule:scalizq}), we must have:
$$\prooftree\Gamma\vdash\ve{t}\type\alpha.(U\to T)\qquad\Gamma\vdash\ve{r}\type U
\justifies\Gamma\vdash\ve{t}\ve{r}\type \alpha.T
\endprooftree$$

By the rule (\ref{rule:scalder}), we must have:
$$\prooftree\Gamma\vdash\ve{t}\type U\to T\qquad\Gamma\vdash\ve{r}\type \beta.U
\justifies\Gamma\vdash\ve{t}\ve{r}\type \beta.T
\endprooftree$$

By combining these two we obtain the rule $\to_E$ presented in Figure~\ref{fig:types}.

\begin{rem}
A good insight into the type system is that, due to the equivalence relation, scalars in types occur only at the top-level and  in the target subtypes of arrow types. This fits very well with the idea that in {\em Lineal} all term constructs are linear, except for abstraction. With this in mind, the syntax of arrow types could have been restricted to $U\to\alpha.V$, or even written as $U\to_\alpha V$ instead. In other words, we could get rid of the type equivalences (\cf~Definition~\ref{def:equiv}) and represent each type equivalence class by just its canonical member. Such a design choice would spare us some lemmas (\cf~Section~\ref{sec:subjectreduction}), but comes at a price:
\begin{iteMize}{$\bullet$}
\item Equivalences between $0.T$ and $0.R$ or $1.T$ and $T$ would still need to be enforced through an equivalence relation or some unelegant case distinctions, at least if we want to maintain them.
\item More generally our aim is to reflect some of the vectorial structure of the terms of {\em Lineal} up at the level of types. In that sense the explicit type equivalences we have given provide a good indication that types have the desired structure.
\end{iteMize}
\end{rem}

\section{Subject reduction}\label{sec:subjectreduction}
The following theorem ensures that typing is preserved by reduction. Having such a property is part of the basic requirements for a type system.

\begin{thm}[Subject Reduction]\label{thm:subjectreduction} For any terms $\ve{t}$, $\ve{t}'$, any context $\Gamma$ and any type $T$, if $\ve{t}\to\ve{t}'$ and $\Gamma\vdash \ve{t}\type T$, then $\Gamma\vdash \ve{t}'\type T$.
\end{thm}

The proof of this theorem is quite long and non-trivial. This is one of the main technical contributions of the paper.

\subsection{Preliminary lemmas}
In order to prove this theorem, we need several auxiliary lemmas giving general properties of our system. We have tried to provide an intuition of every lemma so as to make it easier to follow. Also, we divided them in four groups, reflecting the nature of their statements. Standard and straightforward lemmas are left without proof to improve readability.

\subsubsection{Lemmas about types}
The lemmas in this section are statements about the properties of the types themselves, \ie~the type equivalence relation.
\medskip

It is not so hard to see that every type is equivalent to unit type multiplied by a scalar. A unit type itself can of course always be multiplied by $1$.

\begin{lem}[$\alpha$ unit]\label{lem:alphaunit}
For each type $T$ there exists a unit type $U$ and a scalar $\alpha$ such that $T\equiv\alpha.U$.\qed
\end{lem}

This first lemma should not be misinterpreted however: this does not mean to say that any scalar appearing within a type can be factored out of the type. For example even a simple unit type $X\to\alpha.X$ is not equivalent to $\alpha.(X\to X)$.

The following just says that when two types are equivalent, then the outer left scalars are the same:

\begin{lem}[Unit does not add scalars]\label{lem:unitdoesntaddscalar}
For any unit types $U$, $U'$ and scalars $\alpha,\beta$, if $\alpha.U\equiv\beta.U'$ then, $\alpha=\beta$ and if $\alpha\neq 0$, then $U\equiv U'$.
\end{lem}
\noindent{\em Informal proof.} Following the grammar of unit types, neither $U$ nor $U'$ could contain scalars in this head form but only in the right hand side of an arrow. However, no equivalence rule lets it come out from the right of the arrow and get to the head-form, so if $\alpha.U\equiv\beta.U'$ that means $\alpha=\beta=0$ or $U\equiv U'$ and $\alpha=\beta$. \qed

\subsubsection{Relation between types}
We define a family of relations between types inspired by \cite[def.~4.2.1]{Barendregt92}:

\begin{defi}\label{def:order} For any types $T, R$ any context $\Gamma$ and any term $\ve t$ such that
$$\prooftree\Gamma\vdash\ve t\type T 
\Justifies\Gamma\vdash\ve t\type R
\endprooftree$$
\begin{enumerate}[(1)]
 \item if $X\notin\FV(\Gamma)$, write $T\succ_{X,\Gamma}^{\ve t} R$ if
  \begin{iteMize}{$\bullet$}
   \item either $R\equiv\forall X.T$
   \item or $T\equiv\forall X.S$ and $R\equiv S[U/X]$ for some $U$ and $S$.
  \end{iteMize}
 \item if $\V$ is a set of type variables such that $\V\cap\FV(\Gamma)=\emptyset$, we define $\succeq_{\V,\Gamma}^{\ve t}$ inductively by
  \begin{iteMize}{$\bullet$}
   \item If $X\in \V$ and $T\succ_{X,\Gamma}^{\ve t} R$, then $T\succeq_{\{X\},\Gamma}^{\ve t} R$.
   \item If $\V_1,\V_2\subseteq\V$, $T\succeq_{\V_1,\Gamma}^{\ve t} R$ and $R\succeq_{\V_2,\Gamma}^{\ve t} S$, then $T\succeq_{\V_1\cup\V_2,\Gamma}^{\ve t} S$.
   \item If $T\equiv R$, then $T\succeq_{\V,\Gamma}^{\ve t}R$.
  \end{iteMize}
\end{enumerate}
\end{defi}

\begin{exa}
 Let the following be a valid derivation.
$$\prooftree
  \prooftree
    \prooftree
      \prooftree\Gamma\vdash\ve t\type T\qquad X\notin\FV(\Gamma)
      \justifies\Gamma\vdash\ve t\type \forall X.T
      \using\forall_I
      \endprooftree
    \justifies\Gamma\vdash\ve t\type T[U/X]
    \using\forall_E
    \endprooftree
    \qquad Y\notin\FV(\Gamma)
  \justifies\Gamma\vdash\ve t\type\forall Y.T[U/X]
  \using\forall_I
  \endprooftree
  \qquad
  \forall Y.T[U/X]\equiv R
\justifies\Gamma\vdash\ve t\type R
\using\equiv
\endprooftree$$
Then $T\succeq_{\{X,Y\},\Gamma}^{\ve t} R$.
\end{exa}

\noindent Note that this relation is stable under reduction in the following way:

\begin{lem}[$\succeq$-stability]\label{lem:subjectreductionofrelation}
 For any types $T,R$, set of type variables $\V$, terms $\ve t$ and $\ve r$ and context $\Gamma$, if $T\succeq_{\V,\Gamma}^{\ve t} R$, $\ve t\to\ve r$ and $\Gamma\vdash\ve r\type T$, then $T\succeq_{\V,\Gamma}^{\ve r} R$.
\end{lem}
\proof It suffices to show this for $\succ_{X,\Gamma}^{\ve t}$, with $X\in\V$. Observe that since $T\succ_{X,\Gamma}^{\ve t}R$, then $X\notin\FV(\Gamma)$. We only have to prove that $\Gamma\vdash\ve r\type R$ is derivable from $\Gamma\vdash\ve r\type T$. We proceed now by cases:
  \begin{iteMize}{$\bullet$}
   \item $R\equiv\forall X.T$, then using rules $\forall_I$ and $\equiv$, we can deduce $\Gamma\vdash\ve r\type R$.
   \item $T\equiv \forall X.S$ and $R\equiv S[U/X]$, then using rules $\forall_E$ and $\equiv$, we can deduce $\Gamma\vdash\ve r\type R$.\qed
  \end{iteMize}

\noindent The following lemma states that scalars do not interfere with the relation.

\begin{lem}[Scalars keep order]\label{lem:scalarskeeporder} For any types $T$, $R$, any scalar $\alpha$, context $\Gamma$, set of variable types $V$ and term $\ve t$, if $T\succeq_{\V,\Gamma}^{\ve t} R$ then $\alpha.T\succeq_{\V,\Gamma}^{\alpha.\ve t} \alpha.R$. \qed
\end{lem}
The following lemma states that if two arrow types are ordered, then they are equivalent up to some substitutions.

\begin{lem}[Arrows comparison]\label{lem:arrowscomp} For any types $U, V$, $T, R$, context $\Gamma$, set of type variables $\V$ and term $\ve t$, if
$V\to R\succeq_{\V,\Gamma}^{\ve t} U\to T$, then there exist unit types $\vec{W}$ and variables $\vec{X}\in\V$ such that 
$$U\to T\equiv(V\to R)[\vec{W}/\vec{X}]$$
\end{lem}
\proof \cf~Appendix \ref{proof:arrowscomp}.\qed

\subsubsection{Classic lemmas}
The lemmas in this section are the classic ones, which appear in most subject reduction proofs.
  
Proving subject reduction means proving that each reduction rule preserves the type. The way to do this is to go in the opposite direction to the reduction rule, \ie~to study the reduct so as to understand where it may come from, thereby decomposing the redex in its basic constituents. Generation lemmas accomplish that purpose.

We need five generation lemmas: the classical ones, one for applications (Lemma~\ref{lem:gen-app}) and one for abstractions (Lemma~\ref{lem:gen-abs}); and three new ones for the algebraic rules, one for products by scalars different than $0$ (Lemma~\ref{lem:gen-scalar}) other for product by $0$ (Lemma~\ref{lem:gen-scalar0}) and one for sums (Lemma~\ref{lem:gen-sum}). All of them follow by induction on the typing derivation.

\begin{lem}[Generation lemma (app)]\label{lem:gen-app} For any terms $\ve{t}, \ve{r}$, any type $T$, any scalar $\gamma$ and any context $\Gamma$, if $\Gamma\vdash\ve{t}\ve{r}\type \gamma.T$, then there exist $\alpha, \beta, U, \V$ with $\gamma.R\succeq_{\V,\Gamma}^{\ve t\ve r} \gamma.T$ such that $\Gamma\vdash\ve{r}\type\alpha.U$ and $\Gamma\vdash\ve{t}\type\beta.(U\to R)$ with $\alpha\times\beta=\gamma$. \qed
\end{lem}

\begin{rem}
 Consider the following example: Let $\Gamma\vdash\ve t\type U\to R$ and $\Gamma\vdash\ve r\type U$. Then $\Gamma\vdash\ve t\ve r\type R$ and if $R\succeq_{\V,\Gamma}^{\ve t\ve r} T$, then $\Gamma\vdash\ve t\ve r\type T$. Note that in general, $\ve t$ does not have type $U\to T$. As a counterexample note that $x:X\to X\vdash x:X\to X$ but $x:X\to X\not\vdash x:X\to \forall X.X$.
\end{rem}

\begin{lem}[Generation lemma (abs)]\label{lem:gen-abs}
For any term $\ve{t}$, any type $T$ and any context $\Gamma$, if $\Gamma\vdash\lambda{x}\,\ve{t}\type T$ then there exist $ U$, $R$ and $\V$ such that $\Gamma,x\type U\vdash\ve{t}\type R$ and $U\to R\succeq_{\V,\Gamma}^{\lambda x\,\ve t} T$.
\end{lem}
\proof \cf~Appendix \ref{proof:gen-abs}.\qed

\begin{lem}[Generation lemma (sc)]\label{lem:gen-scalar} For any scalar $\alpha\neq 0$, any context $\Gamma$, any term $\ve{t}$ and any type $T$, if $\Gamma\vdash\alpha.\ve{t}\type\alpha.T$, then $\Gamma\vdash\ve{t}\type T$.\qed
\end{lem}

\begin{lem}[Generation lemma (sc-0)]\label{lem:gen-scalar0} For any context $\Gamma$, any term $\ve{t}$ and  any type $T$, if $\Gamma\vdash 0.\ve{t}\type T$, then there exists $R$ such that $\Gamma\vdash\ve{t}\type R$. \qed
\end{lem}

\begin{lem}[Generation lemma (sum)]\label{lem:gen-sum} For any terms $\ve{t}$, $\ve{r}$, any scalar $\alpha$, any unit type $U$ and any context $\Gamma$, if $\Gamma\vdash\ve{t}+\ve{r}\type\alpha.U$, then there exist $\delta, \gamma\in\Sc$ such that $\Gamma\vdash\ve{t}\type\delta.U$ and $\Gamma\vdash\ve{r}\type\gamma.U$ with $\delta+\gamma=\alpha$.\qed
\end{lem}
\begin{rem} Note that the assumption of the previous lemma could be weakened as $\Gamma\vdash\ve{t}+\ve{r}\type T$, since by Lemma~\ref{lem:alphaunit}, any type $T$ can be written as $\alpha.U$ for some scalar $\alpha$ and unit type $U$.
\end{rem}

The following lemma is quite standard in proofs of subject reduction for System $F$-like systems and can be found for example in \cite[prop.~4.1.19]{Barendregt92} and \cite[prop.~8.2 and 8.5]{Krivine90}. It ensures that when substituting types for type variables or terms for term variables, in an adequate manner, then the type derived remains valid.

\begin{lem}[Substitution lemma]\label{lem:substitution} For any term $\ve{t}$, any base term $\ve{b}$, any types $T$, $\vec{U}$ and any context $\Gamma$,
\begin{enumerate}[\em(1)]
 \item\label{it:substype} If $\Gamma\vdash\ve{t}\type T$ then $\Gamma[\vec{U}/\vec{X}]\vdash\ve{t}\type T[\vec{U}/\vec{X}]$.
 \item\label{it:substerm} If $\Gamma,{x}\type U\vdash\ve{t}\type T$ and $\Gamma\vdash\ve{b}\type U$ then $\Gamma\vdash\ve{t}[\ve{b}/x]\type T$. \qed
\end{enumerate}
\end{lem}

\subsubsection{Lemmas about the scalars}
This section contains the lemmas which make statements about the relative behaviour of the scalars within terms and within types.
For example, scalars appearing in the terms are reflected within the types also. This is formalized in the following lemma, which is proved by induction on the typing derivation.

\begin{lem}[Scaling unit]\label{lem:scaleunit} For any term $\ve t$, scalar $\alpha$, type $T$ and context $\Gamma$, if
$\Gamma\vdash\alpha.\ve{t}\type T$ then there exists $U$ and $\gamma\in\Sc$ such that $T\equiv\alpha.\gamma.U$.\qed
\end{lem}

Base terms are typed by unit types or its equivalent (proof by induction on the typing derivation).

\begin{lem}[Base terms in unit]\label{lem:baseterms} For any base term $\ve{b}$, context $\Gamma$ and type $T$, if $\Gamma\vdash\ve{b}\type T$ then there exists $U\textrm{ such that }T\equiv U$.\qed
\end{lem}

By $ax_\tzero$, it is easy to see that $\ve{0}$ has type $\tzero$, but also by using equivalences between types we have that $\forall X.\tzero$ is equivalent to $\tzero$. Moreover, for any $T$, $\Gamma$ and $\V$ with $\V\cap\FV(\Gamma) = \emptyset$, such that $\tzero\succeq_{\V,\Gamma}^{\ve 0} T$, we have $T\equiv\tzero$. 
Then we can state the following lemma (proof by induction on the typing derivation).

\begin{lem}[Type for $\ve{0}$]\label{lem:type0} For any context $\Gamma$ and type $T$,
if $\Gamma\vdash \ve{0}\type T$ then $T\equiv\tzero$.\qed
\end{lem}

The following theorem is an important one. It says that our \scalar\ type system is polymorphic only in the unit types but not in the general types in the sense that even if it is possible to derive two types for the same term, the outermost-leftmost scalar (\ie~the scalar in the head position) must remain the same. Its proof is not trivial, as it uses several of the previously formulated lemmas.

\begin{thm}[Uniqueness of scalars]\label{thm:uniqscalar} For any term $\ve t$, any context $\Gamma$, any scalars $\alpha$, $\beta$ and any unit types $U$, $V$, if $\Gamma\vdash\ve{t}\type\alpha.U$ and $\Gamma\vdash\ve{t}\type\beta.V$, then $\alpha=\beta$.
\end{thm}
\proof
Structural induction over $\ve{t}$.
	\begin{enumerate}[(1)]
	 \item $\ve{t}=\ve{0}$. Then by Lemmas~\ref{lem:type0} and \ref{lem:unitdoesntaddscalar}, $\alpha=\beta=0$.
	 
	 \item $\ve{t}=x$ or $\ve t=\lambda x\,\ve t'$. Then by Lemmas~\ref{lem:baseterms} and \ref{lem:unitdoesntaddscalar}, $\alpha=\beta=1$.
	 
	 \item $\ve{t}=\gamma.\ve{t}'$. Then by Lemma~\ref{lem:scaleunit}, there exist $\sigma,\delta,U',V'$, such that $\alpha.U\equiv\gamma.\sigma.U'$ and $\beta.V\equiv\gamma.\delta.V'$. If $\gamma=0$, then $\gamma\times\sigma=\gamma\times\delta=0$ and then by Lemma~\ref{lem:unitdoesntaddscalar}, $\alpha=\beta=0$. If $\gamma\neq 0$, then by Lemma~\ref{lem:gen-scalar}, $\Gamma\vdash\ve{t}'\type\sigma.U'$ and $\Gamma\vdash\ve{t}'\type\delta.V'$, so by the induction hypothesis $\sigma=\delta$. Note that, by Lemma~\ref{lem:unitdoesntaddscalar}, $\alpha=\gamma\times\sigma$ and $\beta=\gamma\times\delta$, so $\alpha=\gamma\times\sigma=\gamma\times\delta=\beta$.

	 \item $\ve{t}=\ve{t}_1+\ve{t}_2$. Then by Lemma~\ref{lem:gen-sum}, there exist $\gamma_1,\gamma_2$ such that $\Gamma\vdash\ve{t}_1\type\gamma_1.U$ and $\Gamma\vdash\ve{t}_2\type\gamma_2.U$ with $\gamma_1+\gamma_2=\alpha$; and also by the same lemma, there exist $\delta_1,\delta_2$ such that $\Gamma\vdash\ve{t}_1\type\delta_1.V$ and $\Gamma\vdash\ve{t}_2\type\delta_2.V$ with $\delta_1+\delta_2=\beta$. Then by the induction hypothesis $\gamma_1=\delta_1$ and $\gamma_2=\delta_2$, so  $\alpha=\gamma_1+\gamma_2=\delta_1+\delta_2=\beta$.

	 \item $\ve{t}=\ve{t}_1\ve{t}_2$. Then by Lemma~\ref{lem:gen-app}, there exist $\gamma_1,\gamma_2,W,\V$ and $\alpha.T\succeq_{\V,\Gamma}^{\ve t_1\ve t_2} \alpha.U$ such that $\Gamma\vdash\ve{t}_1\type\gamma_1.(W\to T)$ and $\Gamma\vdash\ve{t}_2\type\gamma_2.W$ with $\gamma_1\times\gamma_2=\alpha$; and also by the same lemma, there exist $\delta_1,\delta_2,W',\V'$ and $\beta.R\succeq_{\V',\Gamma}^{\ve t_1\ve t_2}\beta.V$ such that $\Gamma\vdash\ve{t}_1\type\delta_1.(W'\to R)$ and $\Gamma\vdash\ve{t}_2\type\delta_2.W'$ with $\delta_1\times\delta_2=\beta$. Then by the induction hypothesis $\gamma_1=\delta_1$ and $\gamma_2=\delta_2$, so $\alpha=\gamma_1\times\gamma_2=\delta_1\times\delta_2=\beta$.\qed
	\end{enumerate}
From this theorem, the uniqueness of $\tzero$ follows, in the sense that no term can have type $\tzero$ and some other type $T$ which is not equivalent to $\tzero$.

\begin{cor}[Uniqueness of $\tzero$]\label{cor:uniqbot} For any term $\ve t$ and any context $\Gamma$,
if $\Gamma\vdash\ve{t}\type\tzero$ then for each $T\not\equiv\tzero$, $\Gamma\not\vdash\ve{t}\type T$.
\end{cor}
\proof Assume $\Gamma\vdash\ve{t}\type T$, then by Lemma~\ref{lem:alphaunit}, $T\equiv\alpha.U$. Since $\Gamma\vdash\ve{t}\type\tzero\equiv 0.U$ we obtain by Theorem~\ref{thm:uniqscalar} that $\alpha=0$.\qed

Since $\ve{0}$ has type $\tzero$ which is equivalent to $0.U$ for any $U$, $\ve{0}$ can still act as argument for an abstraction or even be applied to another term. In either case the result is a term of type $\tzero$:

\begin{lem}[Linearity of $\ve{0}$]\label{lem:0linearity} For any term $\ve t$, any context $\Gamma$ and any type $T$,
\begin{multicols}{2}
\begin{enumerate}[\em(1)]
 \item If $\Gamma\vdash\ve{0}\ve{t}\type T$ then $T\equiv\tzero$.
 \item If $\Gamma\vdash\ve{t}\ve{0}\type T$ then $T\equiv\tzero$.
\end{enumerate}
\end{multicols}
\end{lem}
\proof\hfill
\begin{enumerate}[(1)]
  \item\label{ite:1} Let $\Gamma\vdash\ve{0}\ve{t}\type T$. By Lemma~\ref{lem:alphaunit}, $T\equiv\gamma.U$. Moreover, by Lemma~\ref{lem:gen-app}, there exist $\alpha,\beta,U',\V$ and $\gamma.R\succeq_{\V,\Gamma}^{\ve 0\ve t} \gamma.U$ such that $\Gamma\vdash\ve{0}\type\beta.(U'\to R)$ and $\Gamma\vdash\ve{t}\type\alpha.U'$ with $\gamma=\alpha\times\beta$.
  Hence, by Corollary~\ref{cor:uniqbot}, $\beta.(U'\to R)\equiv\tzero\equiv 0.U$, so by Lemma~\ref{lem:unitdoesntaddscalar}, $\beta=0$, then $\gamma=\alpha\times 0=0$, so $T\equiv\gamma.U\equiv\tzero$.
  \item Analogous to (\ref{ite:1}).\qed
 \end{enumerate}

\subsubsection{Subject reduction cases}
The following three lemmas are in fact cases of subject reduction, however, they are also necessary as lemmas in subsequent proofs.

\begin{lem}[Product]\label{lem:prod} For any term $\ve t$, any scalars $\alpha$ and $\beta$ and any context $\Gamma$,
 if $\Gamma\vdash\alpha.(\beta.\ve{t})\type T$ then $\Gamma\vdash(\alpha\times\beta).\ve{t}\type T$.
\end{lem}
\proof By Lemma~\ref{lem:scaleunit}, there exist $U$ and $\gamma\in\Sc$ such that $T\equiv\alpha.\gamma.U$. We proceed now by cases:
\begin{desCription}
  \item\noindent{\hskip-12 pt\bf $\alpha\neq 0$ and $\beta\neq 0$}:\ By Lemma~\ref{lem:gen-scalar}, $\Gamma\vdash\beta.\ve{t}\type\gamma.U$. Moreover, by Lemma~\ref{lem:scaleunit} again, there exist $U'$ and $\gamma'$ such that $\gamma.U\equiv\beta.\gamma'.U'$. So, by Lemma~\ref{lem:gen-scalar}, $\Gamma\vdash\ve{t}\type\gamma'.U'$, from which, using the rule $s_I$ one can derive $\Gamma\vdash(\alpha\times\beta).\ve{t}\type(\alpha\times\beta).\gamma'.U'$.
  Note that $(\alpha\times\beta).\gamma'.U'\equiv\alpha.\beta.\gamma'.U'\equiv\alpha.\gamma.U\equiv T$.

  \item\noindent{\hskip-12 pt\bf $\alpha\neq 0$ and $\beta=0$}:\ By Lemma~\ref{lem:gen-scalar}, $\Gamma\vdash 0.\ve t\type\gamma.U$. Moreover, by Lemma~\ref{lem:gen-scalar0}, there exists $R$ such that $\Gamma\vdash\ve t\type R$, from which, using the rule $s_I$, one can derive $\Gamma\vdash(\alpha\times 0).\ve t\type(\alpha\times 0).R=0.R$. Note that by Lemma~\ref{lem:alphaunit}, there exists $\delta$ and $V$ such that $R\equiv\delta.V$, so $0.R\equiv 0.V\equiv\tzero$. Since $(\alpha\times 0).\ve t=0.\ve t$, by Corollary~\ref{cor:uniqbot}, $\gamma=0$, and note that $0.U\equiv\tzero$, so $0.R\equiv\tzero\equiv 0.U=(\alpha\times 0).U\equiv\alpha.0.U\equiv T$.

  \item\noindent{\hskip-12 pt\bf $\alpha=0$}:\ By Lemma~\ref{lem:gen-scalar0}, there exists $R$ such that $\Gamma\vdash\beta.\ve t\type R$. Then by Lemma~\ref{lem:gen-scalar} or \ref{lem:gen-scalar0}, depending if $\beta=0$ or not, there exists $S$ such that $\Gamma\vdash\ve t\type S$, from which, using the rule $s_I$, one can derive $\Gamma\vdash 0.\ve t\type 0.S$.
  Note that $0.S\equiv\tzero\equiv 0.\gamma.U\equiv T$. Also note that $0.\ve t=(0\times\beta).\ve t$.\qed
\end{desCription}

\begin{lem}[Distributivity]\label{lem:distrib} For any terms $\ve t$ and $\ve r$, any scalar $\alpha$, any context $\Gamma$ and any type $T$, if $\Gamma\vdash\alpha.(\ve{t}+\ve{r})\type T$ then $\Gamma\vdash\alpha.\ve{t}+\alpha.\ve{r}\type T$.
\end{lem}
\proof Let $\Gamma\vdash\alpha.(\ve{t}+\ve{r})\type T$. By Lemma~\ref{lem:scaleunit}, there exist $\alpha, R$ such that $T\equiv\alpha.R$. We proceed now by cases:
\begin{desCription}
  \item\noindent{\hskip-12 pt\bf $\alpha\neq 0$}:\ By Lemma~\ref{lem:gen-scalar}, $\Gamma\vdash\ve{t}+\ve{r}\type R$. By Lemma~\ref{lem:alphaunit}, there exist $\sigma,U$ such that $R\equiv \sigma.U$. So by Lemma~\ref{lem:gen-sum} there exist $\delta$ and $\gamma$ such that $\Gamma\vdash\ve{t}\type\delta.U$ and $\Gamma\vdash\ve{r}\type\gamma.U$ with $\delta+\gamma=\sigma$. Then by rules $s_I$ and $\equiv$, $\Gamma\vdash\alpha.\ve{t}\type(\alpha\times\delta).U$ and $\Gamma\vdash\alpha.\ve{r}\type(\alpha\times\gamma).U$, from which using the rule $+_I$ one can derive $\Gamma\vdash\alpha.\ve{t}+\alpha.\ve{r}\type(\alpha\times\delta+\alpha\times\gamma).U$. Note that $(\alpha\times\delta+\alpha\times\gamma).U\equiv\alpha.\sigma.U\equiv\alpha.R\equiv T$.

  \item\noindent{\hskip-12 pt\bf $\alpha=0$}:\ By Lemma~\ref{lem:gen-scalar0}, there exists $S$ such that $T\equiv 0.S\equiv\tzero$ with $\Gamma\vdash\ve t+\ve r\type S$. In addition, by Lemma~\ref{lem:alphaunit}, there exist $\delta, V$ such that $S\equiv\delta.V$. Then by Lemma~\ref{lem:gen-sum}, there exist $\sigma,\varsigma$ such that $\Gamma\vdash\ve t\type \sigma.V$ and $\Gamma\vdash\ve r\type \varsigma.V$. By rules $s_I$ and $\equiv$, $\Gamma\vdash 0.\ve t\type 0.V$ and $\Gamma\vdash 0.\ve r\type 0.V$, from which using the rule $+_I$ one can derive $\Gamma\vdash 0.\ve t+0.\ve r\type 0.V$. Note that $0.V\equiv\tzero\equiv T$.\qed
\end{desCription}

\begin{lem}[Factorisation]\label{lem:factorisation} For any term $\ve t$, scalars $\alpha$ and $\beta$, type $T$ and context $\Gamma$, if $\Gamma\vdash\alpha.\ve{t}+\beta.\ve{t}\type T$ then $\Gamma\vdash(\alpha+\beta).\ve{t}\type T$.
\end{lem}
\proof
Let $\Gamma\vdash\alpha.\ve{t}+\beta.\ve{t}\type T$. By Lemma~\ref{lem:alphaunit}, there exist $\sigma,V$ such that $T\equiv\sigma.V$. So, by Lemma~\ref{lem:gen-sum}, there exist $\delta, \gamma\in\Sc$ such that $\Gamma\vdash\alpha.\ve{t}\type\delta.V$ and $\Gamma\vdash\beta.\ve{t}\type\gamma.V$ with $\delta+\gamma=\sigma$.
Then by Lemma~\ref{lem:scaleunit}, there exist $\phi$, $\varphi$, $U$ and $U'$ such that $\delta.V\equiv\alpha.\phi.U$ and $\gamma.V\equiv\beta.\varphi.U'$.
So, by Lemma~\ref{lem:unitdoesntaddscalar}, $\delta=\alpha\times\phi$ and $\gamma=\beta\times\varphi$. We proceed now by cases:

\begin{desCription}
 \item\noindent{\hskip-12 pt\bf $\alpha=0$}:\ Then $\delta=0$ and so $\gamma=\sigma$. Thus $\Gamma\vdash\beta.\ve t\type T$. Note that $\beta=\alpha+\beta$.
 \item\noindent{\hskip-12 pt\bf $\alpha\neq 0$, $\beta=0$}:\ Analogous to the previous case.
 \item\noindent{\hskip-12 pt\bf $\alpha,\beta\neq 0$, $\delta=0$}:\ Then $\gamma=\sigma$, so $\Gamma\vdash\beta.\ve t\type T\equiv\beta.\varphi.U'$, then by Lemma~\ref{lem:gen-scalar}, $\Gamma\vdash\ve t:\varphi.U'$. In addition, $\Gamma\vdash\alpha.\ve t\type 0.V\equiv\tzero\equiv\alpha.\tzero$, so by Lemma~\ref{lem:gen-scalar}, $\Gamma\vdash\ve t\type\tzero$. Then by Corollary~\ref{cor:uniqbot}, $\varphi.U'\equiv\tzero$, so by definition of $\equiv$ and Lemma~\ref{lem:unitdoesntaddscalar}, $\varphi=0$ and then $\gamma=0$, so $\sigma=0$ and $T\equiv\tzero$. Using the rule $s_I$ one can derive $\Gamma\vdash(\alpha+\beta).\ve t\type(\alpha+\beta).\tzero\equiv\tzero\equiv T$.
 \item\noindent{\hskip-12 pt\bf $\alpha,\beta\neq 0$, $\gamma=0$}:\ Analogous to the previous case.
 \item\noindent{\hskip-12 pt\bf $\alpha, \beta, \gamma,\delta\neq 0$}:\ Then $\phi,\varphi\neq 0$, so by Lemma~\ref{lem:unitdoesntaddscalar}, $V\equiv U\equiv U'$.
  Then $\Gamma\vdash\alpha.\ve{t}\type\alpha.\phi.V$ and $\Gamma\vdash\beta.\ve{t}\type\beta.\varphi.V$. Hence by Lemma~\ref{lem:gen-scalar}, $\Gamma\vdash\ve{t}\type\phi.V$ and $\Gamma\vdash\ve{t}\type\varphi.V$.  Then by Theorem~\ref{thm:uniqscalar}, $\phi=\varphi$ and then by the rule $s_I$, one can derive $\Gamma\vdash(\alpha+\beta).\ve{t}\type(\alpha+\beta).\phi.V$. Note that $(\alpha+\beta).\phi.V\equiv((\alpha+\beta)\times\phi).V=(\alpha\times\phi+\beta\times\varphi).V=(\delta+\gamma).V=\sigma.V\equiv T$.\qed
\end{desCription}

\subsection{Subject reduction proof}
Now we are able to prove the subject reduction property (Theorem~\ref{thm:subjectreduction}).

\proof We proceed by checking that every reduction rule preserves the type. We give two cases as example, the full proof can be found in Appendix~\ref{proof:subjectreduction}.

\begin{desCription}
  \item\noindent{\hskip-12 pt\bf rule ${(\ve{t}+\ve{r})\ve{u}}\to{\ve{t}\ve{u}+\ve{r}\ve{u}}$}:\ Let $\Gamma\vdash(\ve{t}+\ve{r})\ve{u}\type T\equiv 1.T$. Then, by Lemma~\ref{lem:gen-app}, there exist $\alpha, \beta, U,\V$ and $T'\succeq_{\V,\Gamma}^{(\ve t+\ve r)\ve u} T$ such that $\Gamma\vdash\ve{u}\type\alpha.U$ and $\Gamma\vdash\ve{t}+\ve{r}\type\beta.(U\to T')\equiv 1.\beta.(U\to T')$ with $\alpha\times\beta=1$. Then by Lemma~\ref{lem:gen-sum}, there exist $\delta$ and $\gamma$ such that $\Gamma\vdash\ve{t}\type\delta.(U\to T')$ and $\Gamma\vdash\ve{r}\type\gamma.(U\to T')$ with $\delta+\gamma=\beta$. Then by the rule $\to_E$, $\Gamma\vdash\ve{t}\ve{u}\type(\delta\times\alpha).T'$ and $\Gamma\vdash\ve{r}\ve{u}\type(\gamma\times\alpha).T'$. 
  Using the rule $+_I$, one can derive $\Gamma\vdash\ve{t}\ve{u}+\ve{r}\ve{u}\type (\delta\times\alpha+\gamma\times\alpha).T'=1.T'\equiv T'$ and by Lemma~\ref{lem:subjectreductionofrelation}, $T'\succ_{\V,\Gamma}^{\ve t\ve u+\ve r\ve u} T$.

  \item\noindent{\hskip-12 pt\bf rule ${(\lambda x\,\ve{t})\ve{b}}\to{\ve{t}[\ve{b}/ x]}$}:\ Let $\Gamma\vdash(\lambda x\,\ve{t})\ve{b}\type T$. By the rule $\equiv$, $\Gamma\vdash(\lambda x\,\ve{t})\ve{b}\type 1.T$, so by Lemma~\ref{lem:gen-app}, there exist $\alpha,\beta, U,\V, T'\succeq_{\V,\Gamma}^{(\lambda x\,\ve t)\ve b} T$ such that $\Gamma\vdash\lambda x\,\ve{t}\type \beta.(U\to T')$ and $\Gamma\vdash\ve{b}\type \alpha.U$ with $\alpha\times\beta=1$. Since $\ve{b}$ is a base term, by Lemmas~\ref{lem:baseterms} and \ref{lem:unitdoesntaddscalar}, $\alpha=1$ and so $\beta=1$. Then by Lemma~\ref{lem:gen-abs}, there exist $V$, $R$ and $\V'$ with $V\to R\succeq_{\V',\Gamma}^{\lambda x\,\ve t}U\to T'$. Then by Lemma~\ref{lem:arrowscomp}, there exist $\vec{W}$ and $\vec{X}\in\V'$ such that $U\equiv V[\vec{W}/\vec{X}]$, $T'\equiv R[\vec{W}/\vec{X}]$ and since $\vec{X}\notin\FV(\Gamma)$, $\Gamma=\Gamma[\vec{W}/\vec{X}]$. So, by Lemma~\ref{lem:substitution}(\ref{it:substype}), $\Gamma,x\type U\vdash\ve{t}\type T'$. Thus, by Lemma~\ref{lem:substitution}(\ref{it:substerm}), $\Gamma\vdash\ve{t}[\ve{b}/x]\type T'$, from which, by Lemma~\ref{lem:subjectreductionofrelation}, one obtain $\Gamma\vdash\ve{t}[\ve{b}/x]\type T$.\qed
\end{desCription}

\section{Strong Normalisation, simplified reduction rules and Confluence}\label{sec:strongnormalisation}
The \scalar\ type system is now proved to have the strong normalisation property, \ie~every typable term is strongly normalising, it cannot reduce forever. In order to show this we first set up another type system, which simply `forgets' the scalars. Hence this simpler type system is just a straightforward extension of System $F$, called here $\lasf$ (Definition \ref{def:lasf}). In the literature surrounding not {\em Lineal} but its cousin, the algebraic $\lambda$-calculus, one finds such a System $F$ in \cite{EhrhardLICS10}, which extends the simply typed algebraic $\lambda$-calculus of \cite{VauxRTA07,VauxMSCS09} -- our $\lasf$ is very similar. Secondly we prove strong normalisation for it (Theorem~\ref{thm:strongnormalisationlasf}). Thirdly we show that every term which has a type in \scalar\ has a type in $\lasf$ (Lemma~\ref{lem:corresp}), which entails strong normalisation in \scalar\ (Theorem~\ref{thm:snscalar}).

This strong normalisation proof constitutes the second main technical contribution of the paper. The confluence of a simplified version of {\em Lineal} is derived from it.
\medskip

In this section we use the following notation: $\Tsf$ is the set of types of $\lasf$ (denoted by $A,B,C$). $\Lambda$ is the set of terms of \emph{Lineal}. $\Gamma\Vdash\ve{t}\type A$ says that it is possible to derive the type $A\in\Tsf$ for the term $\ve{t}\in\Lambda$ in the context $\Gamma$ under the typing rules of $\lasf$. We just use $\vdash$ for \scalar. In addition, we use $\mathrm{Name}\rulesf$ to distinguish the names of the typing rules in $\lasf$ from those of \scalar.

\begin{defi}\label{def:lasf} The type grammar of $\lasf$ is the following:
$$A,B,C=X~|~A\to B~|~\forall X.A$$
The typing rules of $\lasf$ are those of System $F$ plus the following rules:
$$\prooftree
	\justifies {\Gamma \Vdash \ve{0}\type{A}}
	\using ax_0\rulesf
\endprooftree
\qquad\qquad
\prooftree{\Gamma\Vdash\ve{t}\type{A} \qquad \Gamma\Vdash\ve{r}\type{A}}
	\justifies{\Gamma\Vdash\ve{t}+\ve{r}\type{A}}
	\using +_I\rulesf
\endprooftree
\qquad\qquad
\prooftree{\Gamma\Vdash\ve{t}\type{A}}
	\justifies{\Gamma\Vdash \alpha.\ve{t}\type{A}}
	\using s_I\rulesf
\endprooftree$$
\end{defi}

\begin{exa}\label{ex:lasfremovescalars}
 In \scalar\ it is possible to derive $\vdash 3.\lambda x\,(x+x)\type 3.(X\to 2.X)$, while in $\lasf$ we have $\vdash 3.\lambda x\,(x+x)\type X\to X$.
\end{exa}

In order to prove strong normalisation we extend the proof for $\lambda 2$. The standard method was invented by Tait \cite{TaitJSL67} for simply typed $\lambda$-calculus and generalized to System $F$ by Girard \cite{GirardPhDThesis}. Our presentation follows \cite[sec.~4.3]{Barendregt92} (who uses an analogous to Krivine's saturated sets \cite[thm.~8.9]{Krivine90}). The following definitions are taken from this reference -- with slight modifications to handle the extra $\lasf$ rules.

The strong normalisation property entails that every term is strongly normalising, so first we define the set of strongly normalising terms. 

\begin{defi}\label{def:SN}
$\SN = \{\ve{t}\in\Lambda\ |\ \ve{t}$ is strongly normalising$ \}$.
\end{defi}

The notion of closure is often captured by the notion of saturated set. We use the notation $\vec{\ve t}=\ve t_1,\dots,\ve t_n$ with $n\geq 0$. Also $\ve r\vec{\ve t}=\ve r\ve t_1\dots\ve t_n$ where if $n=0$ it is just $\ve r$.

\begin{defi}{$\ $}\label{def:SAT}
\begin{enumerate}[(1)]
 \item A subset $X\subseteq \SN$ is called \emph{saturated} if
\begin{enumerate}[(a)]
 \item $\ve{0}\in X$;
 \item for each $x$ and $\vec{\ve{t}}\in \SN$, $x\vec{\ve{t}}\in X$;
 \item if $\ve b\in\SN$ and $\ve{t}[\ve{b}/x]\vec{\ve{r}}\in X$ then $(\lambda x\,\ve{t})\ve{b}\vec{\ve{r}}\in X$;
 \item if (for each $i\in I$, $\ve{t}_i\vec{\ve{r}}\in X$) then $(\sum_{i\in I}\ve{t}_i)\vec{\ve{r}}\in X$;
 \item if (for each $i\in I$, $\ve{u}\ve{t}_i\vec{\ve r}\in X)$ then $\ve{u}(\sum_{i\in I}\ve{t}_i)\vec{\ve{r}}\in X$;
 \item for each $\alpha\in\Sc$, $\ve{t}\in X$ if and only if $\alpha.\ve{t}\in X$;
 \item $\alpha.\ve{t}_1\ve t_2\dots\ve{t}_n\in X$ if and only if $\ve{t}_1\ve t_2\dots(\alpha.\ve{t}_k)\dots\ve{t}_n\in X\ (1\leq k\leq n)$;
 \item for each $\vec{\ve{t}}\in \SN$, $\ve{0}\vec{\ve{t}}\in X$;
 \item for each $\ve{t},\vec{\ve{u}}\in \SN$, $\ve{t}\ve{0}\vec{\ve{u}}\in X$.
\end{enumerate}
where $I$ is a finite set of indices.
 \item $\SAT = \{X\subseteq\Lambda\ |\ X \textrm{ is saturated}\}$
\end{enumerate}
\end{defi}

\noindent The basic idea is to prove that types correspond to saturated sets. In order to achieve this, we define a valuation from types to $\SAT$ (in fact, from type variables to $\SAT$ and then, we define a set in $\SAT$ by using such a valuation).

\begin{defi}{$\ $}\label{def:valSAT}
 \begin{enumerate}[(1)]
  \item A \emph{valuation} in $\SAT$ is a map $\xi\type \mathbb{V}\to \SAT$, where $\mathbb{V}$ is the set of type variables.
  \item For any $A, B\subseteq\Lambda$, we define $A\Rightarrow B=\{\ve t\in\Lambda~|~$for each $\ve r\in A$, $\ve t\ve r\in B\}$.
  \item Given a valuation $\xi$ in $\SAT$, we define for every $T\in \mathbb{T}(\lasf)$ a set $\val{T}_{\xi}\subseteq\Lambda$ as follows:
\begin{eqnarray*}
 \val{X}_{\xi} & = & \xi(X),\textrm{ where }X\in\mathbb{V}\\
 \val{A\to B}_{\xi} & = & \val{A}_{\xi}\Rightarrow\val{B}_{\xi}\\
 \val{\forall X.A}_{\xi}  & = & \bigcap_{Y\in \SAT}\val{A}_{\xi(X:=Y)}
\end{eqnarray*}
 \end{enumerate}
\end{defi}

\begin{lem}\label{lem:TtoSAT}\hfill
 	\begin{enumerate}[\em(1)]
 		 \item $\SN\in \SAT$,
		 \item If $A, B\in \SAT$ then $A\Rightarrow B\in \SAT$,
		 \item Let $\{A_i\}_{i\in I}$ be a collection of members of $\SAT$, $\bigcap_{i\in I} A_i\in \SAT$,
		 \item Given a valuation $\xi$ in $\SAT$ and $A$ in $\Tsf$, then $\val{A}_{\xi}\in \SAT$.
	\end{enumerate}
\end{lem}
\proof \cf~Appendix \ref{proof:TtoSAT}.\qed

Just like in Definition \ref{def:valSAT}, we define another valuation, this time from term variables to base terms. We use it to check what happens when we change every free variable of a term for any other base term.
The basic idea is the following: we define $\rho,\xi\vDash\ve{t}\type A$ to be the property of changing every free term variable in $\ve{t}$ for another term with the help of the valuation $\rho$ (a base term, since term variables only run over base terms) and still having the resulting term in the set $\val{A}_\xi$. So, we define $\Gamma\vDash\ve{t}\type A$ to be the same property, when the property holds for every pair in $\Gamma$ and for every valuations $\rho$ and $\xi$.

This is formalised in the following definition (Definition \ref{def:valuation}) and with this definition, we prove that if a term has a type in a valid context, then the property above holds (Theorem~\ref{thm:soundness}), which yields the strong normalisation theorem (Theorem~\ref{thm:strongnormalisationlasf}) via the concept of \emph{saturated} set (because \emph{saturated} sets are subsets of $\SN$).

\begin{defi}\label{def:valuation}{$\ $}
\begin{iteMize}{$\bullet$}
 \item A \emph{valuation} in $\Lambda_b$ is a map $\rho\type V\to\Lambda_b$, where $V$ is the set of term variables and $\Lambda_b = \{\ve{b}\in\Lambda\ |\ \ve{b}\textrm{ is a base term}\}$.
 \item Let $\rho$ be a valuation in $\Lambda_b$. Then
$\val{\ve{t}}_{\rho} = \ve{t}[x_1:=\rho(x_1),\dots,x_n:=\rho(x_n)],$
where $\vec{x}=x_1,\dots,x_n$ is the set of free variables in $\ve{t}$.
 \item Let $\rho$ be a valuation in $\Lambda_b$ and $\xi$ a valuation in $\SAT$. Then
\begin{iteMize}{$-$}
 \item $\rho,\xi$ \emph{satisfies} $\ve{t}\type A$, notation $\rho,\xi\vDash\ve{t}\type A$, if and only if $\val{\ve{t}}_{\rho}\in\val{A}_\xi$.
 \item $\rho,\xi\vDash\Gamma$ if and only if $\rho,\xi\vDash x\type A$ for all $x\type A$ in $\Gamma$
 \item $\Gamma\vDash\ve{t}\type A$ if and only if for every $\rho,\xi$, $\rho,\xi\vDash\Gamma$ implies $\rho,\xi\vDash\ve{t}\type A.$
\end{iteMize}
\end{iteMize}
\end{defi}

\begin{thm}[Soundness]\label{thm:soundness}
If $\Gamma\Vdash\ve{t}\type A$ then $\Gamma\vDash\ve{t}\type A$.
\end{thm}
\proof We proceed by induction on the derivation of $\Gamma\Vdash\ve{t}\type T$. We show one case as an example. The full proof is in Appendix~\ref{proof:soundness}.

\noindent\parbox{5.3cm}{
\prooftree\Gamma\Vdash\ve{t}\type A\to B\qquad \Gamma\Vdash\ve{r}\type A
  \justifies\Gamma\Vdash\ve{t}\ve{r}\type B
  \using \to_E\rulesf
  \endprooftree}\hspace{0.5cm}\parbox{9.4cm}{
  By the induction hypothesis, $\Gamma\vDash\ve{t}\type A\to B$ and $\Gamma\vDash\ve{r}\type A$. Assume $\rho,\xi\vDash\Gamma$ in order to show $\rho,\xi\vDash\ve{t}\ve{r}\type B$. Then $\rho,\xi\vDash\ve{t}\type A\to B$, \ie~$\val{\ve{t}}_\rho\in\val{A\to B}_\xi=\val{A}_\xi\Rightarrow\val{B}_\xi$ and $\val{\ve{r}}_\rho\in\val{A}_\xi$. Then $\val{\ve{t}\ve{r}}_\rho=\val{\ve{t}}_\rho\ \val{\ve{r}}_\rho\in\val{B}_\xi$, so $\rho,\xi\vDash\ve{t}\ve{r}\type B$.\qed}

\begin{thm}[Strong normalisation for $\lasf$]\label{thm:strongnormalisationlasf}
If $\Gamma\Vdash\ve{t}\type A$, then $\ve{t}$ is strongly normalising.
\end{thm}
\proof
Let $\Gamma\Vdash\ve{t}\type A$. Then by Theorem~\ref{thm:soundness}, $\Gamma\vDash\ve{t}\type A$. Define $\rho_0(x) = x$ for all $x$ and let $\xi$ be a valuation in \SAT. Then $\rho_0,\xi\vDash\Gamma$ (\ie~for all $(x\type B)\in\Gamma$, $\rho_0,\xi\vDash x\type B$ since $x\in\val{B}_{\xi}$ holds because $\val{B}_{\xi}$ is saturated). Therefore $\rho_0,\xi\vDash\ve{t}\type A$, hence $\ve{t}=\val{\ve{t}}_{\rho_0}\in\val{A}_{\xi}\subseteq \SN$.\qed

It is possible to map every type from \scalar\ to a type in $\lasf$ as follows.

\begin{defi}\label{def:map}
Let $(\cdot)\mapsf$ be the following mapping from types in \scalar\ to \Tsf:
$$\bullet~X\mapsf = X \hspace{0.7cm}
\bullet~(\forall X.T)\mapsf = \forall X.T\mapsf \hspace{0.7cm}
\bullet~(U\to T)\mapsf = U\mapsf\to T\mapsf \hspace{0.7cm}
\bullet~(\alpha.T)\mapsf = T\mapsf \hspace{0.7cm}
\bullet~\tzero\mapsf = A$$
where $A$ is an arbitrary fixed type in $\Tsf$.
\end{defi}

We also extend this definition to contexts: $\Gamma\mapsf = \{ (x\type T\mapsf)\ |\ (x\type T)\in\Gamma \}$.

The following lemma ensures that if it is possible to give a type to a term in \scalar\ then it is possible to give to the term the mapped type in $\lasf$. Example \ref{ex:lasfremovescalars} may help to see this.

\begin{lem}[Correspondence with $\lasf$]\label{lem:corresp}
If $\Gamma\vdash\ve{t}\type T$ then $\Gamma\mapsf\Vdash\ve{t}\type T\mapsf$.
\end{lem}
\proof We proceed by induction on the derivation of $\Gamma\vdash\ve{t}\type T$. We show one case as example. The full proof is in Appendix~\ref{proof:corresp}.

\noindent\parbox{4.8cm}{
	$\prooftree\Gamma\vdash\ve{t}\type\alpha.T\qquad\Gamma\vdash\ve{r}\type\beta.T
	\justifies\Gamma\vdash\ve{t}+\ve{r}\type(\alpha+\beta).T
	\using +_I
	\endprooftree$
	}\hspace{0.5cm}\parbox{9.9cm}{
	By the induction hypothesis $\Gamma\mapsf\Vdash\ve{t}\type T\mapsf$ and $\Gamma\mapsf\Vdash\ve{r}\type T\mapsf$, so by the rule $+_I\rulesf$, $\Gamma\mapsf\Vdash\ve{t}+\ve{r}\type T\mapsf = ((\alpha+\beta).T)\mapsf$.\qed}

Strong normalisation arises as a consequence of strong normalisation for $\lasf$ and the above lemma.

\begin{thm}[Strong normalisation]\label{thm:snscalar}
If $\Gamma\vdash\ve{t}\type T$ then $\ve{t}$ is strongly normalising.
\end{thm}
\proof
By Lemma~\ref{lem:corresp}, $\Gamma\mapsf\Vdash\ve{t}\type T\mapsf$, then by Theorem~\ref{thm:strongnormalisationlasf},  $\ve{t}$ is strongly normalising.\qed

\medskip

Taking up again the example of Section~\ref{sec:language}, terms like $\ve{Y}$ are simply not allowed in this typed setting, since Theorem \ref{thm:snscalar} ensures that all the typable terms have a normal form. So we do not have infinities and hence the intuitive reasons for having restrictions $(*)$ on the {\em Factorisation rules} of the linear-algebraic calculus (\cf~the reduction rules in Section~\ref{sec:language}) have now vanished. If we drop them, the example becomes as follows.
\begin{exa}
Consider some arbitrary typable and hence normalising term $\ve{t}$. Then $\alpha.\ve{t}-\alpha.\ve{t}$ can be reduced by a factorisation rule into $(\alpha-\alpha).\ve{t}$. This reduces in one step to $\ve{0}$, without the need to reduce $\ve{t}$. 
\end{exa}
It turns out that, in general, for typable terms we can indeed drop the restrictions (*) and (**) without breaking the confluence of \emph{Lineal}. These restrictions were there only due to the impossibility of checking for the normalisation property in the untyped setting. 
Having this property, the confluence of the system follows naturally.
\begin{thm}[Confluence]\label{thm:confluence}
 Let $\ve t$ be a term of \emph{Lineal}, as it appears in Figure~\ref{fig:lineal}, but without restrictions (*) and (**). If $\ve t$ is typable in \scalar, $\ve t\to^*\ve{u}$ and $\ve t\to^*\ve r$, then there exists a term $\ve t'$ such that $\ve{u}\to^*\ve t'$ and $\ve r\to^*\ve t'$.
\end{thm}
\proof First let us introduce some notation. Let $\to_\beta$ be a $\beta$-reduction and $\to_a$ any reduction from Figure~\ref{fig:lineal} but the $\beta$-reduction, without restrictions (*) and (**). Moreover, $\to_\beta^*$ and $\to_a^*$ are the application of zero or more of such rules, as usual.

The proof follows in several steps.
\begin{enumerate}[(1)]
  \item First we prove local confluence for the algebraic fragment, \ie~if $\ve t\to_a\ve{u}$ and $\ve t\to_a\ve r$, then there exists a term $\ve t'$ such that $\ve{u}\to_a^*\ve t'$ and $\ve r\to_a^*\ve t'$.
  \item Then we prove local confluence for the $\beta$-reduction, \ie~if $\ve t\to_\beta\ve{u}$ and $\ve t\to_\beta\ve r$, then there exists a term $\ve t'$ such that $\ve{u}\to_\beta^*\ve t'$ and $\ve r\to_\beta^*\ve t'$.
  \item\label{cor:confluence:comm} Finally, we prove that algebraic rules and $\beta$-reduction commutes, \ie~if $\ve t\to_a\ve{u}$ and $\ve t\to_\beta\ve r$, then there exists a term $\ve t'$ such that $\ve{u}\to^*\ve t'$ and $\ve r\to^*\ve t'$, where $\to^*$ is a reduction sequence of zero or more steps involving any rules.
\end{enumerate}
This prove the local confluence of the system. The well known Newman's lemma states that local confluence plus strong normalisation implies confluence (\cf~for example~\cite[thm.~1.2.1]{Terese03}).

This is how the proofs of the steps look like.
\begin{enumerate}[(1)]
 \item Valiron did a semi-automatized proof in the interactive theorem prover Coq~\cite{ManualCOQ}. The interested reader can find the proof in~\cite{BenCoqProofVectorial}. An easier-to-read explanation of it appears in \cite{ArrighiDiazcaroValironDCM11}.
 \item The local confluence of the $\beta$-reduction follows from~\cite{DoughertyIC92} and a trivial extension of the confluence of lambda-calculus.
 \item This proof goes by structural induction. \cf~Appendix~\ref{proof:confluence}. \qed
\end{enumerate}
Note that the proofs of subject reduction (Theorem~\ref{thm:subjectreduction}) and strong normalisation (Theorem~\ref{thm:snscalar}) have been done in the general case, without restrictions (*) and (**), so they are still valid for the simplified calculus.

Having dropped restrictions (*) and (**) is an important simplification of the linear-algebraic $\lambda$-calculus, which becomes really just an oriented version of the axioms of vector spaces \cite{ArrighiDowekWRLA04} together with the call-by-value strategy (\ie~restriction (***) remains, of course, to make all functions remain linear in their arguments, in the sense of linear-algebra).

\section{Barycentric \texorpdfstring{$\lambda$}{lambda}-calculus}\label{sec:disc}

By slightly modifying our system, the \scalar\ type system may be used in order to specialize {\em Lineal} into a higher-order barycentric calculus. In order to illustrate this point, let us consider the following type judgement, which can be obtained from \scalar:
$$f::=\lambda x\,((x(\frac{1}{2}.(\true+\false)))(\frac{1}{4}.\true+\frac{3}{4}.\false))\type\mathbb{B}\to\mathbb{B};$$
where $\mathbb{B}$ stands for $\forall X.X\to X\to X$. Note that the type $\mathbb{B}$ has $\true$, $\false$ and linear combinations of them with scalars summing to one, as members. In this example the type system provides a guarantee that the function is barycentric and that if it receives a barycentric argument, it preserves this property. For example, if we apply such a function to $\frac{1}{2}.(\true+\false)$ we obtain:
$$f(\frac{1}{2}.(\true+\false))\longrightarrow^*\frac{3}{8}.\true+\frac{5}{8}.\false.$$

Although this seems feasible, we do not develop a full-blown barycentric higher-order $\lambda$-calculus and associated properties in this paper. We just show that the \scalar\ type system accomplishes part of the job by checking for the barycentric property, \ie~checking that the normal form of a term has amplitudes summing to one. A barycentric $\lambda$-calculus fragment of the algebraic $\lambda$-calculus has already been studied for its own sake \cite{TassonTLCA09}, however in this work the calculus was endowed with a simple type system, not one that would recognize barycentric terms amongst other terms.

To this end let us define a type system with the rules and grammar of \scalar, but where the valid types are the classic ones (\ie~types exempt of any scalar, which we have referred to as~\Tsf\ in Definition~\ref{def:lasf}), whilst all the other types are just intermediate types:
\begin{defi}\label{def:probtypesystem}
  We define the type system $\B$ for the barycentric calculus to be the \scalar\ type system with the following restrictions:
  \begin{iteMize}{$\bullet$}
	\item $\Sc=\mathbb{R}$,
	\item Contexts are sets of tuples $(x\type A)$, with $A\in\Tsf$,
	\item Type variables run over \Tsf~instead of unit types, \ie~the rule $\forall_E$ accepts only $A\in\Tsf$,
	\item The final sequent has type in $\Tsf$.
  \end{iteMize}
 Note that the type derivation may have, as intermediate sequents, sequents that do not have their type in~$\Tsf$ but in \scalar.
\end{defi}
\begin{exa}\label{ex:typinginB}
 The sequent $y\type X\vdash 2.(\lambda x\,\frac{1}{2}.x)y\type X$ is valid in $\B$, even if intermediate sequents do not have their types restricted~to~$\Tsf$:
$$\prooftree
   \prooftree
     \prooftree
        \prooftree
	  \prooftree
	    \prooftree
	    \justifies y\type X,x\type X\vdash x\type X
	    \using ax
	    \endprooftree
	  \justifies y\type X,x\type X\vdash \tfrac{1}{2}.x\type \tfrac{1}{2}.X
	  \using s_I
	  \endprooftree
	\justifies y\type X\vdash \lambda x\,\tfrac{1}{2}.x\type X\to\tfrac{1}{2}.X
	\using\to_I
	\endprooftree
	\prooftree
	\justifies y\type X\vdash y\type X
	\using ax
	\endprooftree
     \justifies y\type X\vdash(\lambda x\,\tfrac{1}{2}.x)y\type \tfrac{1}{2}.X
     \using\to_E
     \endprooftree
   \justifies y\type X\vdash 2.(\lambda x\,\tfrac{1}{2}.x)y\type 2.\tfrac{1}{2}.X
   \using s_I
   \endprooftree
   \qquad 2.\tfrac{1}{2}.X\equiv X
\justifies y\type X\vdash 2.(\lambda x\,\tfrac{1}{2}.x)y\type X
\using\equiv
\endprooftree$$
\end{exa}

In order to show that this type system does the job, let us define the weight function:
\begin{defi}
  Let $\omega:\Lambda\to\mathbb{R}$ be a function defined inductively by:
  $$\begin{array}{l@{\hspace{1cm}}l@{\hspace{1cm}}r}
  \omega(\ve{0}) = 0	& \omega(\ve{b})= 1 & \omega(\ve{t}_1+\ve{t}_2)= \omega(\ve{t}_1)+\omega(\ve{t}_2)\\
  \multicolumn{2}{l}{\omega(\ve{t}_1\ve{t}_2)= \omega(\ve{t}_1)\times\omega(\ve{t}_2)}	& \omega(\alpha.\ve{t}) = \alpha\times\omega(\ve{t})\\
  \end{array}$$
  where $\ve{b}$ is a base term.
\end{defi}

\begin{exa}\label{ex:omeganotenough}
Continuing Example~\ref{ex:typinginB}, note that $2.(\lambda x\,\frac{1}{2}.x)y\to^* y$.
The weight function cannot check the barycentric property on the first term (indeed $\omega(2.(\lambda x\,\frac{1}{2}.x)y)=2$), even when this term reduces to a barycentric term. For this matters, we use the type system $\B$ to check this property, as stated by the following theorem.
\end{exa}

\begin{thm}[Normal-form of terms in \texorpdfstring{$\mathcal{B}$}{B} have weight $1$]\label{thm:weight1}
Let $\Gamma\vdash\ve{t}\type A$ with $A\in\Tsf$, then $\omega(\nf{\ve{t}})=1$.
\end{thm}
We need the following three lemmas in order to prove this theorem. The first lemma says that an application in normal form can only happen when a variable is in head position.

\begin{lem}\label{lem:R2}
If $\ve{t}_1\ve{t}_2$ is in normal form, then $\ve{t}_1=x\vec{\ve{r}}$.
\end{lem}
\proof Structural induction on $\ve{t}_1$. \cf~Appendix~\ref{proof:R2}.\qed

The following lemma says that if there is an application in normal form (which by the previous lemma has to have the form $x\vec{\ve{r}}$), then it cannot be typed in the type system $\mathcal{B}$ with a type having a scalar at the right of an arrow, the only possible scalar have to be in the head position of the type.
\begin{lem}\label{lem:R3}
$\Gamma\vdash x\vec{\ve{r}}\type T$ and $x\vec{\ve{r}}$ is in normal form, then there exist $A\in\Tsf$ and $\alpha\in\Sc$ such that $T\equiv\alpha.A$.
\end{lem}
\proof Induction on the derivation of $\Gamma\vdash x\vec{\ve{r}}\type T$. \cf~Appendix~\ref{proof:R3}.\qed

The last lemma shows that scalars cannot be removed by subsumption.
\begin{lem}\label{lem:scalarsnotaddedbysubsumption}
 Let $A\in\Tsf$. For any $T,\V,\Gamma,\ve t$, if $\alpha.T\succeq_{\V,\Gamma}^{\ve t}\alpha.A$, then $T\equiv B$ with $B\in\Tsf$.
\end{lem}
\proof Case analysis. \cf~Appendix~\ref{proof:scalarsnotaddedbysubsumption} \qed

Using these lemmas, the proof of Theorem~\ref{thm:weight1} goes as follows.
\proof Instead, we prove the more general case: If $\Gamma\vdash\ve{t}\type\alpha.A$ then $\omega(\nf{\ve{t}})=\alpha$, by structural induction on $\nf{\ve{t}}$. We take $\Gamma\vdash\nf{\ve{t}}\type\alpha.A$, which is true by Theorem~\ref{thm:subjectreduction}.
We show two cases as an example. The full proof is in Appendix~\ref{proof:weight1}.

\begin{iteMize}{$\bullet$}
\item Case $\nf{\ve{t}}=\gamma.\ve{t}'$. Then $\omega(\nf{\ve{t}})=\gamma\times\omega(\ve{t}')$. By Lemma~\ref{lem:scaleunit}, there exist $U$ and $\delta$ such that $\alpha.A\equiv\gamma.\delta.U$. Note that $\Tsf$ is included in the set of unit types, so by Lemma~\ref{lem:unitdoesntaddscalar}, $\alpha=\gamma\times\delta$. We consider two cases:
\begin{desCription}
  \item\noindent{\hskip-12 pt\bf $\alpha=0$}:\ Then either $\gamma=0$, and so $\omega(\gamma.\ve{t}')=0\times\omega(\ve{t}')=0$, or $\gamma\neq 0$ but $\delta=0$, and so by Lemma~\ref{lem:gen-scalar}, $\Gamma\vdash\ve{t}'\type 0.U\equiv 0.A$, so by the induction hypothesis $\omega(\ve{t}')=0$, and then $\omega(\gamma.\ve{t}')=\gamma\times 0=0$.
  \item\noindent{\hskip-12 pt\bf $\alpha\neq 0$}:\ Then $A\equiv U$, so by Lemma~\ref{lem:gen-scalar}, $\Gamma\vdash\ve{t}'\type\delta.A$. Then by the induction hypothesis $\omega(\ve{t}')=\delta$. Note that $\omega(\nf{\ve{t}})=\gamma\times\omega(\ve{t}')=\gamma\times\delta=\alpha$.
\end{desCription}
\item Case $\nf{\ve{t}}=\ve{t}_1\ve{t}_2$. Then $\omega(\nf{\ve{t}})=\omega(\ve{t}_1)\times\omega(\ve{t}_2)$. By Lemma~\ref{lem:gen-app}, there exist $U, T, \beta, \delta$ and $\V$ such that $\Gamma\vdash\ve{t}_1\type\beta.(U\to T)$ and $\Gamma\vdash\ve{t}_2\type\delta.U$ with $\alpha.T\succeq_{\V,\Gamma}^{\ve t_1\ve t_2}\alpha.A$ and $\beta\times\delta=\alpha$.
  Since $A\in\Tsf$ by assumption, using Lemma~\ref{lem:scalarsnotaddedbysubsumption} we have $T\equiv B$ with $B\in\Tsf$ and since $\ve{t}_1\ve{t}_2$ is in normal form, by Lemma~\ref{lem:R2}, $\ve{t}_1$ is a variable applied to something else, so by Lemma~\ref{lem:R3}, $U\to B\in\Tsf$, which implies that $U\in\Tsf$. Then by the induction hypothesis, $\omega(\ve{t}_1)=\beta$ and $\omega(\ve{t}_2)=\delta$, so $\omega(\nf{\ve{t}})=\omega(\ve{t}_1)\times\omega(\ve{t}_2)=\beta\times\delta=\alpha$.\qed
\end{iteMize}
 
\begin{rems}\hfill
\begin{iteMize}{$\bullet$}
 \item As first proved in \cite[prop.~2]{ArrighiDowekRTA08}, closed normal terms have form 
$$\sum\limits_{i=1}^n \alpha_i.\lambda x\,\ve{t}_i+\sum\limits_{j=1}^m \lambda x\,\ve{u}_j$$
Thus the Theorem~\ref{thm:weight1} entails that $\sum_{i=1}^n \alpha_i+m=1$.
Hence the type system $\mathcal{B}$, an easy variation of the \scalar\ type system, checks for the barycentric property, \ie~it checks that a given term reduces to a barycentric distribution of terms.
 \item A bit more elaborated example than the one given in Example~\ref{ex:omeganotenough}. It is easy to prove that $$z:A,w:A\vdash(2.\lambda x\,\lambda y\,(\frac{1}{4}.x+\frac{1}{4}.y))zw\type A.$$ But note that $\omega((2.\lambda x\,\lambda y\,(\frac{1}{4}.x+\frac{1}{4}.y))zw)=2$, even when $(2.\lambda x\,\lambda y\,(\frac{1}{4}.x+\frac{1}{4}.y))zw\to^*\frac{1}{2}.z+\frac{1}{2}.w$, whose weight is equal to one. So, \emph{a priori} this $\omega$ function cannot tell us that this term yields a barycentric term. However the fact that has type $A$ in $\Tsf$, according to the Theorem~\ref{thm:weight1}, anticipates this result.
  \item One might think that unit types are just as good as barycentric types $\B$ for the sake of obtaining Theorem \ref{thm:weight1}, via a combination of Subject-reduction and the uniqueness of scalars property (\cf~Theorem \ref{thm:uniqscalar}). This is not quite the case, here is a counter-example:
$$ x:U\to2.U,\,y:U\vdash x\dfrac{1}{2}.y:U\quad\textrm{ but }\omega(x\dfrac{1}{2}.y)=\dfrac{1}{2}$$
But the more significant difference between unit types and $\B$ is one of composability: the application of a term of unit type to another is not necessarily of unit type; whereas barycentric types on the other hand are preserved under application. Thus terms in $\B$ are not only barycentric; they can also be viewed as barycentric-preserving functions. 
\end{iteMize}
\end{rems}

\section{Summary of contributions}\label{sec:conclusion}

In summary, we have defined a System $F$-like type system for an extension of {\em Lineal}, a $\lambda$-calculus which allows making arbitrary linear combinations of $\lambda$-calculus terms $\alpha.\ve{t}+\beta.\ve{u}$. The \scalar\ type system is fine-grained in that it keeps track of the `amount of a type', \ie~the scalar of the type of a term is the sum of the scalars of the types of the contributing subterms.\\

Our main technical contributions were:
\begin{iteMize}{$\bullet$}
\item A proof of the subject reduction property of the \scalar\ type system (Theorem~\ref{thm:subjectreduction}). This came out after having proven a set of lemmas related to the equivalence relation intrinsic to the types and another set of lemmas explaining how the scalars within the types are related to the scalars within the terms. Once all of the important properties were known, we were able to use them to decompose and recompose any term before and after applying a reduction rule, so as to show that every reduction rule preserves the types.
\item A proof of the strong normalisation property of the \scalar\ type system (Theorem~\ref{thm:snscalar}). The technique used to prove the strong normalisation property was by proving that such property would hold for a simpler system and then to show the correspondence between the two systems. 
\item A proof that under strong normalisation, most of the conditions upon the {\em Lineal} reduction rules can be lifted (\eg~allowing the factorisation not only of closed normal terms but of any term) without jeopardizing confluence, thereby simplifying the {\em Lineal} language.
\item A proof that the \scalar\ type system can be used to check that a term has the barycentric property, \ie~that the amplitudes of its normal form are summing to one.
\end{iteMize}
Arguably a denotational semantics approach might have led to less syntactic proofs of the properties of the type system, sustained by the guiding intuition about an underlying mathematical space. On the other hand, the complexity of the proofs in this paper is largely due to the large number of rules ($16$ rules plus associativity and commutativity of $+$). Moreover the issue of models of (linear-)algebraic $\lambda$-calculus is a challenging, active topic of the current research. We know of the categorical model of simply typed {\em Lineal} \cite{ValironDCM10} and the finiteness space model of simply typed algebraic $\lambda$-calculus \cite{EhrhardMSCS05,TassonTLCA09}. Moreover, even if both calculi simulate each other \cite{DiazcaroPerdrixTassonValironHOR10}, it is not clear whether the translation applies at the level of models. Hence known models are intricate and tend not to cover the set of terms under consideration in this paper. Note also that since the models of untyped $\lambda$-calculus are uncountable, the models of (linear-)algebraic $\lambda$-calculus are likely to be vector spaces of uncountable dimensions. These are fascinating, open questions. \phantom{\cite{LeBotlanRemySIGPLAN03}}

\section*{Acknowledgement}
We would like to thank to Gilles Dowek, Jonathan Grattage, Philippe Jorrand, Simon Perdrix, Barbara Petit, Fr\'ed\'eric Prost and Beno\^{\i}t Valiron for enlightening discussions.
We thank also the anonymous referee for his careful reading, which enabled us to improve the level of rigour of our proofs.

\bibliographystyle{alpha}
\bibliography{biblio}

\appendix\label{appendices}
\section{Omitted proofs}
\subsection{Proof of Lemma~\ref{lem:arrowscomp}}\label{proof:arrowscomp}

A map $(\cdot)^\circ$ from types of \scalar\ to \Tsf is defined by
$$X^\circ = X\qquad
 (U\to T)^\circ = U\to T\qquad
 (\forall X.T)^\circ = T^\circ\qquad
 (\alpha.T)^\circ = \alpha.T^\circ\qquad
 (\tzero)^\circ = \tzero$$

We need three intermediate results.
\begin{enumerate}[(1)]
 \item\label{it:ir0} If $T\equiv R$, then $T^\circ\equiv R^\circ$.
 \item\label{it:ir1} For any types $T$ and $U$, exists $V$ such that $(T[U/X])^\circ\equiv T^\circ[V/X]$
 \item\label{it:ir2} For any types $T$, $R$, set of type variables $\V$, context $\Gamma$ and term $\ve t$, if $T\succeq_{\V,\Gamma}^{\ve t} R$ then there exists $\vec{U}$ and $\vec{X}\in\V$ such that $R^\circ\equiv T^\circ[\vec{U}/\vec{X}]$
\end{enumerate}
\textit{Proofs.}
\begin{enumerate}[(1)]
 \item By induction on the equivalence rules. We only show the basic cases since the inductive step, given by the context where the equivalence is applied, is trivial.
        \begin{desCription}
         \item\noindent{\hskip-12 pt\bf $\alpha.\tzero\equiv\tzero$}:\ $(\alpha.\tzero)^\circ =\alpha.\tzero^\circ=\alpha.\tzero\equiv\tzero=\tzero^\circ$.
	 \item\noindent{\hskip-12 pt\bf $0.T\equiv\tzero$}:\ $(0.T)^\circ=0.T^\circ\equiv\tzero=\tzero^\circ$.
	 \item\noindent{\hskip-12 pt\bf $1.T\equiv T$}:\ $(1.T)^\circ=1.T^\circ\equiv T^\circ$.
	 \item\noindent{\hskip-12 pt\bf $\alpha.(\beta.T)\equiv(\alpha\times\beta).T$}:\ $(\alpha.(\beta.T))^\circ=\alpha.(\beta.T^\circ)\equiv(\alpha\times\beta).T^\circ=((\alpha\times\beta).T)^\circ$.
	 \item\noindent{\hskip-12 pt\bf $\forall X.\alpha.T\equiv\alpha.\forall X.T$}:\ $(\forall X.\alpha.T)^\circ=(\alpha.T)^\circ=\alpha.T^\circ=\alpha.(\forall X.T)^\circ=(\alpha.\forall X.T)^\circ$.
        \end{desCription}
 \item Structural induction on $T$
	\begin{desCription}
	 \item\noindent{\hskip-12 pt\bf $T=X$}:\ then $(X[U/X])^\circ=U^\circ=X[U^\circ/X]=X^\circ[U^\circ/X]$.
	 \item\noindent{\hskip-12 pt\bf $T=Y$}:\ then $(Y[U/X])^\circ=Y=Y^\circ[U/X]$.
	 \item\noindent{\hskip-12 pt\bf $T= V\to R$}:\ then $((V\to R)[U/X])^\circ=(V[U/X]\to R[U/X])^\circ=V[U/X]\to R[U/X]=(V\to R)[U/X]=(V\to R)^\circ[U/X]$.
	 \item\noindent{\hskip-12 pt\bf $T=\forall Y.R$}:\ then $((\forall Y.R)[U/X])^\circ=(\forall Y.R[U/X])^\circ=(R[U/X])^\circ$, which, by the induction hypothesis, is $\equiv$ to $R^\circ[V/X]=(\forall Y.R)^\circ[V/X]$.
	 \item\noindent{\hskip-12 pt\bf $T=\tzero$}:\ analogous to $T\equiv Y$.
	 \item\noindent{\hskip-12 pt\bf $T=\alpha.R$}:\ then $(\alpha.R[U/X])^\circ=\alpha.(R[U/X])^\circ$, which, by the induction hypothesis, is $\equiv$ to $\alpha.(R^\circ)[V/X]=(\alpha.R)^\circ[V/X]$.
	\end{desCription}
 \item It suffices to show this for $T\succ_{X,\Gamma}^{\ve t} R$.
	\begin{desCription}
	 \item\noindent{\hskip-12 pt\bf Case 1}:\ $R\equiv\forall X.T$. Then $R^\circ\equiv T^\circ$.
	 \item\noindent{\hskip-12 pt\bf Case 2}:\ $T\equiv\forall X.S$ and $R\equiv S[U/X]$ then by the intermediate results \ref{it:ir0} and \ref{it:ir1}, one has $R^\circ\equiv(S[U/X])^\circ\equiv S^\circ[V/X]\equiv T^\circ[V/X]$ for some $V$.
	\end{desCription}
\end{enumerate}
Proof of the lemma. $U\to T\equiv(U\to T)^\circ$, by the intermediate result \ref{it:ir2}, this is $\equiv$ to $(V\to R)^\circ[\vec{U}/\vec{X}]=(V\to R)[\vec{U}/\vec{X}]$. \qed

\subsection{Proof of Lemma~\ref{lem:gen-abs}}\label{proof:gen-abs}
Induction on the depth of the derivation.

\begin{enumerate}
\item \parbox{3.4cm}{
  \prooftree\Gamma,x\type U\vdash\ve{t}\type T
  \justifies\Gamma\vdash\lambda x\,\ve{t}\type U\to T
  \using \to_I
  \endprooftree
  }\hspace{0.5cm}\parbox{10cm}{
  This is the trivial case. Note that for each $\V\not\subseteq\FV(\Gamma)$, one has $U\to T\succ_{\V,\Gamma}^{\lambda x\,\ve t}U\to T$.}
\bigskip

\item \parbox{3.5cm}{
  \prooftree\Gamma\vdash\lambda x\,\ve{t}\type \forall X.T
  \justifies\Gamma\vdash\lambda x\,\ve{t}\type T[V/X]
  \using\forall_E
  \endprooftree
  }\hspace{0.5cm}\parbox{9.9cm}{
  By the induction hypothesis there exist $U$, $R$ and $\V$ such that $\Gamma,x\type U\vdash\ve{t}\type R$ and $U\to R\succeq_{\V,\Gamma}^{\lambda x\,\ve t}\forall X.T \succ_{X,\Gamma}^{\lambda x\,\ve t} T[V/X]$. (Observe that since $X$ is bounded in $\forall X.T$, any variable in $FV(\Gamma)$ is distinct from it).}
\bigskip

\item \parbox{4.7cm}{
  \prooftree\Gamma\vdash\lambda x\,\ve{t}\type T\quad X\notin\FV(\Gamma)
  \justifies\Gamma\vdash\lambda x\,\ve{t}\type\forall X.T
  \using\forall_I
  \endprooftree
  }\hspace{0.5cm}\parbox{8.7cm}{
  By the induction hypothesis there exist $U$, $R$ and $\V$ such that $\Gamma,x\type U\vdash\ve{t}\type R$ and $U\to R\succeq_{\V,\Gamma}^{\lambda x\,\ve t} T \succ_{X,\Gamma}^{\lambda x\,\ve t} \forall X.T$.}
\bigskip

\item\parbox{4.1cm}{
  \prooftree\Gamma\vdash\lambda x\,\ve t\type T\qquad T\equiv R
  \justifies\Gamma\vdash\lambda x\,\ve t\type R
  \using\equiv
  \endprooftree
  }\hspace{0.5cm}\parbox{9.3cm}{
  By the induction hypothesis there exist $U$, $S$ and $\V$ such that $\Gamma,x\type U\vdash\ve t\type S$ and $U\to S\succeq_{\V,\Gamma}^{\lambda x\,\ve t} T\equiv R$.\qed}

\end{enumerate}

\subsection{Proof of Theorem~\ref{thm:subjectreduction}}\label{proof:subjectreduction}
We proceed by checking that every reduction rule preserves the type. Let $\ve t\to\ve r$ and $\Gamma\vdash\ve t\type T$. To show that $\Gamma\vdash\ve r\type T$, we proceed by induction on the derivation of the typing judgement.

\begin{desCription}
\item\noindent{\hskip-12 pt\bf \emph{Elementary rules}}:\ \hfill
\begin{desCription}
\item\noindent{\hskip-12 pt\bf rule ${\ve{t}+\ve{0}}\to{\ve{t}}$}:\ Let $\Gamma\vdash\ve{t}+\ve{0}\type T$. By Lemma~\ref{lem:alphaunit}, there exist $U$ and $\alpha$ such that  $T\equiv\alpha.U$, then by Lemma~\ref{lem:gen-sum}, there exist $\delta, \gamma\in\Sc$ such that $\Gamma\vdash\ve{t}\type\delta.U$ and $\Gamma\vdash\ve{0}\type\gamma.U$ with $\delta+\gamma=\alpha$. By Lemma~\ref{lem:type0}, $\gamma.U\equiv\tzero$, so $\gamma=0$ and then $\delta=\alpha$.

\item\noindent{\hskip-12 pt\bf rule ${0.\ve{t}}\to{\ve{0}}$}:\ Let $\Gamma\vdash 0.\ve{t}\type T$. By Lemma~\ref{lem:scaleunit}, there exists $R$ such that $T\equiv 0.R\equiv\tzero$ and by the rule $ax_\tzero$, $\Gamma\vdash\ve{0}\type\tzero$.

\item\noindent{\hskip-12 pt\bf rule ${1.\ve{t}}\to{\ve{t}}$}:\ Let $\Gamma\vdash 1.\ve{t}\type T\equiv 1.T$. By Lemma~\ref{lem:gen-scalar}, $\Gamma\vdash\ve{t}\type T$.

\item\noindent{\hskip-12 pt\bf rule ${\alpha.\ve{0}}\to{\ve{0}}$}:\ Let $\Gamma\vdash\alpha.\ve{0}\type T$. By Lemma~\ref{lem:scaleunit}, there exists $R$ such that $T\equiv\alpha.R$. We proceed now by cases:
\begin{desCription}
\item\noindent{\hskip-12 pt\bf $\alpha\neq 0$}:\ By Lemma~\ref{lem:gen-scalar}, $\Gamma\vdash\ve{0}\type R$. Thus, by Lemma~\ref{lem:type0}, $R\equiv\tzero$ and so $T\equiv\alpha.R\equiv\tzero$.
\item\noindent{\hskip-12 pt\bf $\alpha=0$}:\ By Lemma~\ref{lem:gen-scalar0} there exists $S$ such that $\Gamma\vdash\ve{0}\type S$ and by Lemma~\ref{lem:type0} $S\equiv\tzero$.
\end{desCription}

\item\noindent{\hskip-12 pt\bf rule ${\alpha.(\beta.\ve{t})}\to{(\alpha\times\beta).\ve{t}}$}:\ True by Lemma~\ref{lem:prod}

\item\noindent{\hskip-12 pt\bf rule ${\alpha.(\ve{t}+\ve{r})}\to{\alpha.\ve{t}+\alpha.\ve{r}}$}:\ True by Lemma~\ref{lem:distrib}.
\end{desCription}

\item\noindent{\hskip-12 pt\bf\emph{Factorisation rules}}:\ \hfill
\begin{desCription}
\item\noindent{\hskip-12 pt\bf rule ${\alpha.\ve{t}+\beta.\ve{t}}\to{(\alpha+\beta).\ve{t}}$}:\ True by Lemma~\ref{lem:factorisation}.

\item\noindent{\hskip-12 pt\bf rule ${\alpha.\ve{t}+\ve{t}}\to{(\alpha+1).\ve{t}}$}:\ Let $\Gamma\vdash\alpha.\ve{t}+\ve{t}\type T$. Using the rule $s_I$ one can derive $\Gamma\vdash 1.(\alpha.\ve{t}+\ve{t})\type 1.T$. Then by Lemma~\ref{lem:distrib}, $\Gamma\vdash 1.\alpha.\ve{t}+1.\ve{t}\type 1.T$. Moreover, by Lemma~\ref{lem:alphaunit}, there exist $\beta,U$ such that $T\equiv\beta.U$ and then by Lemma~\ref{lem:gen-sum}, $\Gamma\vdash 1.\alpha.\ve{t}\type\gamma.U$ and $\Gamma\vdash 1.\ve{t}\type\delta.U$ with $\gamma+\delta=\beta$. So, by Lemma~\ref{lem:prod}, $\Gamma\vdash\alpha.\ve{t}\type \gamma.U$. Then using the rule $+_I$ one can derive $\Gamma\vdash \alpha.\ve{t}+1.\ve{t}\type \alpha.U$. We conclude, by Lemma~\ref{lem:factorisation}, with $\Gamma\vdash(\alpha+1).\ve{t}\type \alpha.U\equiv T$.

\item\noindent{\hskip-12 pt\bf rule ${\ve{t}+\ve{t}}\to{(1+1).\ve{t}}$}:\ Let $\Gamma\vdash\ve{t}+\ve{t}\type T$. Then by the rule $s_I$, $\Gamma\vdash 1.(\ve{t}+\ve{t})\type 1.T$. By Lemma~\ref{lem:distrib}, $\Gamma\vdash 1.\ve{t}+1.\ve{t}\type 1.T$ and by Lemma~\ref{lem:factorisation}, $\Gamma\vdash(1+1).\ve{t}\type 1.T\equiv T$.
\end{desCription}

\item\noindent{\hskip-12 pt\bf \emph{Application rules}}:\ \hfill
\begin{desCription}
\item\noindent{\hskip-12 pt\bf rule ${(\ve{t}+\ve{r})\ve{u}}\to{\ve{t}\ve{u}+\ve{r}\ve{u}}$}:\ Given in the main text of the paper.

\item\noindent{\hskip-12 pt\bf rule ${\ve{u}(\ve{t}+\ve{r})}\to{\ve{u}\ve{t}+\ve{u}\ve{r}}$}:\ Analogous to the previous case.

\item\noindent{\hskip-12 pt\bf rule ${(\alpha.\ve{t})\ve{r}}\to{\alpha.\ve{t}\ve{r}}$}:\ Let $\Gamma\vdash(\alpha.\ve{t})\ve{r}\type T\equiv 1.T$. Then by Lemma~\ref{lem:gen-app}, there exist $\gamma,\beta,U,\V$ and $T'\succeq_{\V,\Gamma}^{(\alpha.\ve t)\ve r} T$ such that $\Gamma\vdash\ve{r}\type\gamma.U$ and $\Gamma\vdash\alpha.\ve{t}\type\beta.(U\to T')$ with $\gamma\times\beta=1$. Moreover, by Lemma~\ref{lem:scaleunit}, $\beta.(U\to T')\equiv\alpha.\delta.U'$ then by Lemma~\ref{lem:unitdoesntaddscalar}, $U\to T'\equiv U'$ and $\beta=\alpha\times\delta$ (note that $\beta\neq 0$ because $\gamma\times\beta=1$). So by Lemma~\ref{lem:gen-scalar}, $\Gamma\vdash\ve{t}\type\delta.(U\to T')$. By using the rule $\to_E$ one can derive $\Gamma\vdash\ve{t}\ve{r}\type\delta.\gamma.T'$ from which, using the rule $s_I$ one can deduce $\Gamma\vdash\alpha.\ve{t}\ve{r}\type\alpha.\delta.\gamma.T'$. Note that $\alpha.\delta.\gamma.T'\equiv(\alpha\times\delta\times\gamma).T'=(\beta\times\gamma).T'=1.T'\equiv T'$ and by Lemma~\ref{lem:subjectreductionofrelation}, $T'\succeq_{\V,\Gamma}^{\alpha.\ve{t}\ve{r}} T$.

\item\noindent{\hskip-12 pt\bf rule ${\ve{r}(\alpha.\ve{t})}\to{\alpha.\ve{r}\ve{t}}$}:\ Analogous to the previous case.

\item\noindent{\hskip-12 pt\bf rule ${\ve{0}\ve{t}}\to{\ve{0}}$}:\ True by Lemma~\ref{lem:0linearity} and the rule $ax_\tzero$.

\item\noindent{\hskip-12 pt\bf rule ${\ve{t}\ve{0}}\to{\ve{0}}$}:\ True by Lemma~\ref{lem:0linearity} and the rule $ax_\tzero$.
\end{desCription}

\item\noindent{\hskip-12 pt\bf\emph{Beta reduction}}:\ Given in the main text of the paper.

\item\noindent{\hskip-12 pt\bf\emph{AC equivalences}}:\ \hfill
  \begin{desCription}
	\item\noindent{\hskip-12 pt\bf Commutativity}:\ Let $\Gamma\vdash\ve{t}+\ve{r}\type T$. By Lemma~\ref{lem:alphaunit}, there exist $\alpha, U$ such that $T\equiv\alpha.U$. Then, by Lemma~\ref{lem:gen-sum}, there exist $\delta$ and $\gamma$ such that $\Gamma\vdash\ve{t}\type\delta.U$ and $\Gamma\vdash\ve{r}\type\gamma.U$ with $\delta+\gamma=\alpha$. Then using rules $+_I$ and $\equiv$, one can derive $\Gamma\vdash\ve{r}+\ve{t}\type T$.

	\item\noindent{\hskip-12 pt\bf Associativity}:\ Let $\Gamma\vdash(\ve{t}+\ve{r})+\ve{u}\type T$. By Lemma~\ref{lem:alphaunit}, there exist $\alpha, U$ such that $T\equiv\alpha.U$. Then, by Lemma~\ref{lem:gen-sum}, there exist $\delta$ and $\gamma$ such that $\Gamma\vdash\ve{t}+\ve{r}\type\delta.U$ and $\Gamma\vdash\ve{u}\type\gamma.U$ with $\delta+\gamma=\alpha$. Then, by Lemma~\ref{lem:gen-sum} again, there exist $\delta'$ and $\gamma'$ such that $\Gamma\vdash\ve{t}\type\delta'.U$ and $\Gamma\vdash\ve{r}\type\gamma'.U$ with $\delta'+\gamma'=\delta$. Then with the rule $+_I$ one can deduce $\Gamma\vdash\ve{r}+\ve{u}\type(\gamma'+\gamma).U$ and with the same rule, $\Gamma\vdash\ve{t}+(\ve{r}+\ve{u})\type(\delta'+\gamma'+\gamma).U \equiv T$. The inverse is analogous: if $\Gamma\vdash\ve{t}+(\ve{r}+\ve{u})\type T$ then $\Gamma\vdash(\ve{t}+\ve{r})+\ve{u}\type T$.
  \end{desCription}

\item\noindent{\hskip-12 pt\bf\emph{Contextual rules}}:\ Let $\ve t\to\ve r$ and assume as the induction hypothesis that for any context $\Gamma$ and type $T$, if $\Gamma\vdash\ve t\type T$ then $\Gamma\vdash\ve r\type T$.
  \begin{desCription}
	\item\noindent{\hskip-12 pt\bf $\ve t\ve u\to\ve r\ve u$}:\ Let $\Gamma\vdash\ve t\ve u\type T$. By the rule $\equiv$, $\Gamma\vdash\ve t\ve u\type 1.T$. Then by Lemma~\ref{lem:gen-app}, there exist $\alpha$, $\beta$, $U$, $R$ and $\V$ such that $\Gamma\vdash\ve t\type\alpha.(U\to R)$ and $\Gamma\vdash\ve u\type\beta.U$ with $R\succeq_{\V,\Gamma}^{\ve t\ve u} T$ and $\alpha\times\beta=1$. By the induction hypothesis $\Gamma\vdash\ve r\type\alpha.(U\to R)$, from which, using the rule $\to_E$, one can deduce $\Gamma\vdash\ve r\ve u\type\alpha\times\beta.R$. Note that $\alpha\times\beta.R=1.R\equiv R$ and by Lemma~\ref{lem:subjectreductionofrelation}, $R\succeq_{\V,\Gamma}^{\ve r\ve u} T$.
	
	\item\noindent{\hskip-12 pt\bf $\ve u\ve t\to\ve u\ve r$}:\ Analogous to the previous case.

	\item\noindent{\hskip-12 pt\bf $\ve t+\ve u\to\ve r+\ve u$}:\ Let $\Gamma\vdash\ve t+\ve u\type T$. By Lemma~\ref{lem:alphaunit}, $T\equiv\alpha.U$, so by the rule $\equiv$, $\Gamma\vdash\ve t+\ve u\type \alpha.U$. Then by Lemma~\ref{lem:gen-sum}, $\Gamma\vdash\ve t\type\delta.U$ and $\Gamma\vdash\ve u\type\gamma.U$ with $\delta+\gamma=\alpha$. By the induction hypothesis $\Gamma\vdash\ve r\type\delta.U$, so using the rule $+_I$ one can deduce $\Gamma\vdash\ve r+\ve u\type(\delta+\gamma).U$. Note that $(\delta+\gamma).U=\alpha.U\equiv T$.
	
	\item\noindent{\hskip-12 pt\bf $\ve u+\ve t\to\ve u+\ve r$:}\ Analogous to the previous case.
	
	\item\noindent{\hskip-12 pt\bf $\alpha.\ve t\to\alpha.\ve r$}:\ Let $\Gamma\vdash\alpha.\ve t\type T$. By Lemma~\ref{lem:scaleunit}, there exist $\gamma$ and $U$ such that $T\equiv\alpha.\gamma.U$. We proceed now by cases:
	  \begin{desCription}
	   \item\noindent{\hskip-12 pt\bf $\alpha\neq 0$}:\ By Lemma~\ref{lem:gen-scalar}, $\Gamma\vdash\ve t\type\gamma.U$. Moreover, by the induction hypothesis $\Gamma\vdash\ve r\type\gamma.U$ and using the rule $s_I$ one can derive $\Gamma\vdash\alpha.\ve r\type\alpha.\gamma.U\equiv T$.
	   \item\noindent{\hskip-12 pt\bf $\alpha=0$}:\ By Lemma~\ref{lem:gen-scalar0}, there exists $R$ such that $\Gamma\vdash\ve t\type R$. Moreover, by the induction hypothesis $\Gamma\vdash\ve r\type R$ and using the rule $s_I$ one can derive $\Gamma\vdash\alpha.\ve r\type\alpha.R\equiv\tzero\equiv T$.
	  \end{desCription}
	\item\noindent{\hskip-12 pt\bf $\lambda x\,\ve t\to\lambda x\,\ve r$}:\ Let $\Gamma\vdash\lambda x\,\ve t\type T$. By Lemma~\ref{lem:gen-abs}, there exist $U$, $R$ and $\V$ such that $\Gamma,x\type U\vdash\ve t\type R$ with $U\to R\succeq_{\V,\Gamma}^{\lambda x\,\ve t} T$. Then by the induction hypothesis $\Gamma,x\type U\vdash\ve r\type R$, and using the rule $\to_I$ one can derive $\Gamma\vdash\lambda x\,\ve r\type U\to R$. By Lemma~\ref{lem:subjectreductionofrelation}, $U\to R\succeq_{\V,\Gamma}^{\lambda x\,\ve r} T$. \qed
  \end{desCription}
\end{desCription}

\subsection{Proof of Lemma~\ref{lem:TtoSAT}}\label{proof:TtoSAT}
\begin{enumerate}[(1)]
	\item Obviously $\SN\subseteq \SN$. We need to prove it satisfies each point of the definition of saturation.
	\begin{enumerate}
		\item $\ve{0}\in \SN$.
		\item\label{ite:var-proof:TtoSAT1} for each $x$ and $\vec{\ve t}$, $x\vec{\ve t}\in \SN$.
	 	\item\label{ite:abs-proof:TtoSAT1} Assume $\ve{t}[\ve{b}/x]\vec{\ve{r}}\in \SN$, then the term
		\begin{equation}\label{eq:2}
		 (\lambda x\,\ve{t})\ve{b}\vec{\ve r}
		\end{equation}
		must terminate because $\ve{t}, \ve{b}$ and $\vec{\ve r}$ terminate since they are $\SN$ by assumption ($\ve{t}[\ve{b}/x]$ is a sub-term of a term in $\SN$, hence itself is $\SN$; but then $\ve{t}$ is also $\SN$), After finitely many steps reducing terms in \ref{eq:2} we obtain $(\lambda x\,\ve{t}')\ve{b}'\vec{\ve r'}$ with $\ve{t}\to^*\ve{t}'$, $\ve b\to\ve b'$ and for each $i$, $\ve r_i\to\ve r'_i$. Then the contraction of $(\lambda x\,\ve{t}')\ve{b}'\vec{\ve{r}'}$ gives
		\begin{equation}\label{eq:3}
		 \ve{t}'[\ve{b}'/x]\vec{\ve r'}
		\end{equation}
		This is a reduct of $\ve{t}[\ve{b}/x]\vec{\ve r}$ and since this is $\SN$, also \ref{eq:3} and $(\lambda x\,\ve{t})\ve{b}\vec{\ve{r}}$ are $\SN$.
		\item\label{ite:sumapp-proof:TtoSAT1} First note that if $\ve{t}, \ve{u}\in \SN$, then $\ve{t}+\ve{u}\in \SN$. Now, assume that for each $i\in I$, $\ve{t}_i\vec{\ve{r}}\in \SN$, which implies that $\ve t_i$ and $\vec{\ve r}$ are $\SN$.
	 Also, note that $(\sum_{i\in I}\ve{t}_i)\vec{\ve{r}}\to^*\sum_{i\in I}\ve{t}_i\vec{\ve{r}}$ which is the sum of $\SN$ terms, so is $\SN$. We need to prove that any other reduction starting from  $\sum_{i\in I}\ve{t}_i\vec{\ve{r}}$ is also $\SN$. We proceed by induction on $I$. To simplify the notation, we take $I=\{1,\dots,n\}$ with $n\geq 1$.
		\begin{iteMize}{$\bullet$}
			\item If $I=\{1\}$, then we are done, since for each $i\in I$, $\ve{t}_i\vec{\ve{r}}\in \SN$.
			\item Assume it is true for $I=\{1,\dots,n\}$, that is $(\sum_{i=1}^n\ve t_i)\vec{\ve r}\in \SN$.
			\item Let $I=\{1,\dots,n+1\}$, then we must prove that $(\sum_{i=1}^n\ve t_i+\ve t_{n+1})\vec{\ve r}\in \SN$. We proceed by a case analysis of every possible reductions of this term. Note that any reduction in $\vec{\ve t}$ or $\vec{\ve r}$ is finite since these terms are in $\SN$, so the amount of addends is the same.\\
			{\bf \em Elementary rules:}\ \hfill
				\begin{iteMize}{$-$}
					\item One $\ve t_k=\ve 0$ and the rule $\ve t+\ve 0\to\ve t$ applies. Then the induction hypothesis closes the case.
					\item One $\ve t_k=\alpha.(\ve t_{k_1}+\ve t_{k_2})$ and reduces to $\alpha.\ve t_{k_1}+\alpha.\ve t_{k_2}$, it still can be considered as one addend since $\ve t_k$ is in $\SN$.
					\item In other case, it is just a reduction in one $\ve t_i$ or one of $\vec{\ve r}$.
				\end{iteMize}
				{\bf \em Factorisation rules:}\ This case follows by the induction hypothesis.\\
				{\bf \em Application rules:}\ \hfill
				\begin{iteMize}{$-$}
					\item Again, reductions on any $\ve t_k$ are not considered since these cases are trivial by the strong normalisation of $\ve t_k$.
					\item $(\sum_{i=1}^n\ve t_i+\ve t_{n+1})\vec{\ve r}\to(\sum_{i=1}^n\ve t_i)\vec{\ve r}+(\ve t_{n+1})\vec{\ve r}$, then since by the induction hypothesis $(\sum_{i=1}^n\ve t_i)\vec{\ve r}\in \SN$ and by assumption also $(\ve t_{n+1})\vec{\ve r}\in \SN$, its sum is in $\SN$.
					\item $(\sum_{i=1}^n\ve t_i+\ve t_{n+1})\vec{\ve r}\to(\sum_{i=1}^k\ve t_i)\vec{\ve r}+(\sum_{i=k+1}^{n+1}\ve t_i)\vec{\ve r}$ the induction hypothesis applies to both addends.
					\item Any other case does not involve a sum, so they are either included in the case of reduction of one $\ve t_k$ or one from $\vec{\ve r}$, or in the basic case of the induction. Recall that the terms are considered modulo AC axioms for $+$ so it is not relevant to consider the permutations or different parentheses locations in the big sums.
				\end{iteMize}
				{\bf \em Beta reduction:}\ This is either the basic case or a reduction in one $\ve t_k$ or one of $\phantom{x}\enspace\,\,\vec{\ve r}$.
		\end{iteMize}
		\item for each $i\in I$, $\ve{u}\ve{t}_i\vec{\ve{r}}\in \SN$. Note that $\ve{u}(\sum_{i\in I}\ve{t}_i)\vec{\ve{r}}\to^*\sum_{i\in I}\ve{u}\ve{t}_i\vec{\ve{r}}$ which is the sum of $\SN$ terms. This case is analogous to \ref{ite:sumapp-proof:TtoSAT1}.
		\item\label{ite:scalarofSN-proof:TtoSAT1} if $\ve{t}\in \SN$, then for each $\alpha\in\Sc, \alpha.\ve{t}\in \SN$ and vice-versa.
		\item\label{ite:scalar-proof:TtoSAT1} $\alpha.\ve{t}_1\ve t_2\dots\ve{t}_n\in \SN$ then for each $k$, the term $\ve{t}_1\dots(\alpha.\ve{t}_k)\dots\ve{t}_n$ must terminate because $\vec{\ve{t}}$ terminate since these terms are $\SN$ by assumption, so after finitely many reduction steps reducing $\ve{t}_1\dots(\alpha.\ve{t}_k)\dots\ve{t}_n$ we obtain $\alpha.\ve{u}$, with $\ve{t}_1\dots\ve{t}_n\to^*\ve{u}$. So $\alpha.\ve{u}$ is a reduct of $\alpha.\ve{t}_1\dots\ve{t}_n$ and since this term is $\SN$, for each $k$, $\ve{t}_1\dots(\alpha.\ve{t}_k)\dots\ve{t}_n$ are $\SN$.
		\item\label{ite:0algo-proof:TtoSAT1} $\ve{0}\vec{\ve{t}}\to^*\ve{0}\vec{\ve{t}'}$ and since $\vec{\ve{t}}$ is $\SN$, assume $\vec{\ve{t}'}$ is in normal form, so $\ve{0}\vec{\ve{t}'}$ can only reduce to $\ve{0}$, then $\ve{0}\vec{\ve{t}}\in \SN$.
		\item\label{ite:algo0-proof:TtoSAT1} Consider the term $\ve{t}\ve{0}\vec{\ve{u}}$. The term reduces to $\ve 0\vec{\ve u}$, which then reduces to $\ve 0$. But we could also choose to first reduce $\ve t$ or $\vec{\ve u}$. In any case, the terms $\ve t$ and $\ve u$ do not interact with any other term than with $\ve 0$, and when they do they reach $\ve 0$. This will necessarily happen since $\ve t$ and $\ve u$ are in $\SN$ by assumption. Hence the entire term is in $\SN$.
	\end{enumerate}
	\item Let $A, B\in \SAT$, then $x\in A$ by definition of saturated sets. For each $\ve t\in A\Rightarrow B$, $\ve tx\in B$. Since $B\in \SAT$, then $B\subseteq \SN$, so $\ve tx\in \SN$ and so $\ve t$ is strongly normalising.
	Therefore $A\Rightarrow B\subseteq \SN$.
	Now we need to show $A\Rightarrow B$ is saturated by showing each point in the definition of saturated sets.
	
	\begin{enumerate}[(a)]
		\item By saturation of $B$, for each $\ve{u}\in A$, $\ve{0}\ve{u}\in B$, then $\ve{0}\in A\Rightarrow B$.
		\item Let $\vec{\ve{t}}\in \SN$, we need to show that $x\vec{\ve{t}}\in A\Rightarrow B$, \ie~for each $\ve{u}\in A$, $x\vec{\ve{t}}\ve{u}\in B$, which is true since $A\in \SAT$ implies that $\ve{u}\in \SN$, so $B\in \SAT$ implies that $x\vec{\ve{t}}\ve{u}\in B$.
		\item Let $\ve{t}[\ve{b}/x]\vec{\ve{r}}\in A\Rightarrow B$, then for each $\ve{u}\in A$, $\ve{t}[\ve{b}/x]\vec{\ve{r}}\ve{u}\in B$ and since $B$ is saturated, $(\lambda x\,\ve{t})\ve{b}\vec{\ve{r}}\ve{u}\in B$, so $(\lambda x\,\ve{t})\ve{b}\vec{\ve{r}}\in A\Rightarrow B$.
		\item Let for each $i\in I$, $\ve{t}_i\vec{\ve{r}}\in A\Rightarrow B$, then for each $\ve{u}\in A$ and for each $i\in I$, $\ve{t}_i\vec{\ve{r}}\ve{u}\in B$, then by the saturation of $B$, $(\sum_{i\in I}\ve{t}_i)\vec{\ve{r}}\ve{u}\in B$, so $(\sum_{i\in I}\ve{t}_i)\vec{\ve{r}}\in A\Rightarrow B$.
		\item Let for each $i\in I$, $\ve{u}\ve{t}_i\vec{\ve{r}}\in A\Rightarrow B$, then for each $\ve t'\in A$, $\ve{u}\ve{t}_i\vec{\ve{r}}\ve{t}'\in B$, then by saturation of $B$, $\ve{u}(\sum_{i\in I}\ve{t}_i)\vec{\ve{r}}\ve{t}'\in B$, so $\ve{u}(\sum_{i\in I}\ve{t}_i)\vec{\ve{r}}\in A\Rightarrow B$.
		\item Let $\ve{t}\in A\Rightarrow B$ then for each $\ve{u}\in A$, $\ve{t}\ve{u}\in B$, then by the saturation of $B$, for each $\alpha\in\Sc$, $\alpha.\ve{t}\ve{u}\in B$, then also by the saturation of $B$, $(\alpha.\ve{t})\ve{u}\in B$, so $\alpha.\ve{t}\in A\Rightarrow B$.
		
		Let $\alpha.\ve{t}\in A\Rightarrow B$, then for each $\ve{u}\in A$, $\alpha.\ve{t}\ve{u}\in B$, so by the saturation of $B$, $\alpha.\ve{t}\ve{u}\in B$, and again, by the saturation of $B$, $\ve{t}\ve{u}\in B$, so $\ve{t}\in A\Rightarrow B$.
		\item Let $\alpha.\ve{t}_1\dots\ve{t}_n\in A\Rightarrow B$, then for each $\ve{u}\in A$, $(\alpha.\ve{t}_1\dots\ve{t}_n)\ve{u}\in B$, then by the saturation of $B$, $\alpha.\ve{t}_1\dots\ve{t}_n\ve{u}\in B$, and so, by the saturation of $B$ again, for each $k$, $\ve{t}_1\dots(\alpha.\ve{t}_k)\dots\ve{t}_n\ve{u}\in B$, then for each $k$, $\ve{t}_1\dots(\alpha.\ve{t}_k)\dots\ve{t}_n\in A\Rightarrow B$.
		
		The inverse follows analogously: let $\ve{t}_1\dots(\alpha.\ve{t}_k)\dots\ve{t}_n\in A\Rightarrow B$, then for each $\ve{u}\in A$, $\ve{t}_1\dots(\alpha.\ve{t}_k)\dots\ve{t}_n\ve{u}\in B$, so by the saturation of $B$, one has $\alpha.\ve{t}_1\dots\ve{t}_n\ve{u}\in B$ and then, also by the saturation of $B$, one has $(\alpha.\ve{t}_1\dots\ve{t}_n)\ve{u}\in B$, then $\alpha.\ve{t}_1\dots\ve{t}_n\in A\Rightarrow B$.
		
		\item For each $\ve u\in A$, $\ve u\in \SN$ and then, by the saturation of $B$, for each $\vec{\ve t}\in \SN$, $\ve 0\vec{\ve t}\ve u\in B$. Then $\ve{0}\vec{\ve{t}}\in A\Rightarrow B$.
		
		\item For each $\ve r\in A$, $\ve r\in \SN$ and then, by the saturation of $B$, for each $\ve t,\vec{\ve u}\in \SN$, $\ve t\ve 0\vec{\ve u}\ve r\in B$. Then $\ve t\ve 0\vec{\ve u}\in A\Rightarrow B$.
	\end{enumerate}
	
	\item Let $\{A_i\}_{i\in I}$ be a collection of members of $\SAT$, then for each $i\in I$, $A_i\subseteq \SN$, so $\bigcap_{i\in I}A_i\subseteq \SN$.
	We have to show that $\bigcap_{i\in I}A_i$ is saturated.
	\begin{iteMize}{$\bullet$}
		\item Conditions (a), (b), (h) and (i) follow trivially: all these conditions have the form ``$\ve t\in X$''. Since by the saturation of $A_i$, for each $i\in I, \ve t\in A_i$, then $\ve t\in\bigcap_{i\in I}A_i$.
		\item Conditions (c), (d), (e), (f) and (g) are also straightforward: all these conditions have the form ``If $\ve t$ in $X$, then $\ve r$ in $X$''. Let $\ve t\in\bigcap_{i\in I}A_i$, then for each $i\in I, \ve t\in A_i$ and so, by the saturation of $A_i$, $\ve r\in A_i$, from where one can deduce $\ve r\in\bigcap_{i\in I}A_i$.
	\end{iteMize}
	
	\item By structural induction on $A$.
	\begin{desCription}
		\item\noindent{\hskip-12 pt\bf $A:=X$}:\ Then $\val{A}_\xi=\xi(X)\in \SAT$.
		\item\noindent{\hskip-12 pt\bf $A:=B\to C$}:\ Then $\val{A}_\xi=\val{B}_\xi\Rightarrow\val{C}_\xi$. By the induction hypothesis $\val{B}_\xi$ and $\val{C}_\xi\in \SAT$, then by Lemma~\ref{lem:TtoSAT}(2), $\val{B}_\xi\Rightarrow\val{C}_\xi\in \SAT$.
		\item\noindent{\hskip-12 pt\bf $A:=\forall X.A'$}:\ Then $\val{A}_\xi=\bigcap_{Y\in \SAT}\val{A'}_{\xi(X:=Y)}$. By the induction hypothesis $\forall Y\in \SAT, \val{A'}_{\xi(X:=Y)}\in \SAT$, then by Lemma~\ref{lem:TtoSAT}(3), $\bigcap_{Y\in \SAT}\val{A'}_{\xi(X:=Y)}\in \SAT$. \qed
	\end{desCription}
\end{enumerate}

\subsection{Proof of Theorem~\ref{thm:soundness}}\label{proof:soundness}
We proceed by induction on the derivation of $\Gamma\Vdash\ve{t}\type T$.
\begin{enumerate}
\item\parbox{3.1cm}{
\prooftree
\justifies\Gamma,x\type A\Vdash x\type A
\using ax\rulesf
\endprooftree}\hspace{0.5cm}\parbox{10.3cm}{
Note that if $\rho,\xi\vDash\Gamma,x\type A$, then by definition $\rho,\xi\vDash x\type A$.}
\medskip

\item\parbox{2.2cm}{
\prooftree
\justifies\Gamma\Vdash\ve{0}\type A
\using ax_0\rulesf
\endprooftree}\hspace{0.5cm}\parbox{11.2cm}{
Then for each $\xi$ and $\rho$, by the saturation of $\val{A}_\xi$ one has $\ve{0}\in\val{A}_\xi$. Since $\val{\ve{0}}_\rho = \ve{0}$, then $\rho,\xi\vDash\ve{0}\type A$, and so for every $\Gamma$, $\Gamma\vDash\ve{0}\type A$.}
\bigskip

\item\parbox{5.3cm}{
\prooftree\Gamma\Vdash\ve{t}\type A\to B\qquad \Gamma\Vdash\ve{r}\type A
\justifies\Gamma\Vdash\ve{t}\ve{r}\type B
  \using \to_E\rulesf
\endprooftree}\hspace{0.5cm}\parbox{8.1cm}{Given in the main text of the paper.}
\medskip

\item\parbox{3.5cm}{
\prooftree\Gamma,x\type A\Vdash\ve{t}\type B
\justifies\Gamma\Vdash\lambda x\,\ve{t}\type A\to B
\using\to_I\rulesf
\endprooftree}\hspace{0.5cm}\parbox{9.9cm}{Assume $\rho,\xi\vDash\Gamma$ in order to show $\rho,\xi\vDash\lambda x\,\ve{t}\type A\to B$. That is, we must show $\val{\lambda x\,\ve{t}}_\rho\ve{u}\in\val{B}_\xi$ for all $\ve{u}\in\val{A}_\xi$. Assume $\ve{u}\in\val{A}_\xi$, then $\rho(x:=\ve{u})\vDash\Gamma,x\type A$ and hence by the induction hypothesis $\val{\ve{t}}_{\rho(x:=\ve{u})}\in B$. Since $\val{\lambda x\,\ve{t}}_\rho\ve{u}= (\lambda x\,\ve{t})[\vec{y}:=\rho(\vec{y})]\ve{u}\to_\beta
\ve{t}[\vec{y}:=\rho(\vec{y}),x:=\ve{u}] = \val{\ve{t}}_{\rho(x:=\ve{u})}$, it follows from the saturation of $\val{B}_\xi$ that $\val{\lambda x\,\ve{t}}_\rho\ve{u}\in\val{B}_\xi$.}
\medskip

\item\parbox{3.1cm}{
\prooftree\Gamma\Vdash\ve{t}\type\forall X.A
\justifies\Gamma\Vdash\ve{t}\type A[B/X]
\using\forall_E\rulesf
\endprooftree}\hspace{0.5cm}\parbox{10.3cm}{
Assume $\rho,\xi\vDash\Gamma$ in order to show $\rho,\xi\vDash\ve{t}\type A[B/X]$. By the induction hypothesis $\val{\ve{t}}_\rho\in\val{\forall X.A}_\xi$ and this set is equal to $\bigcap_{Y\in \SAT}\val{A}_{\xi(X:=Y)}$, hence $\val{\ve{t}}_\rho\in\val{A}_{\xi(X:=\val{B}_\xi)}=\val{A[B/X]}_\xi$.}
\medskip

\item\parbox{4.6cm}{
\prooftree\Gamma\Vdash\ve{t}\type A\qquad X\notin \FV(\Gamma)
\justifies\Gamma\Vdash\ve{t}\type\forall X.A
\using\forall_I\rulesf
\endprooftree}\hspace{0.5cm}\parbox{8.8cm}{
Assume $\rho,\xi\vDash\Gamma$ in order to show $\rho,\xi\vDash\ve{t}\type \forall X.A$. Since $X\notin \FV(\Gamma)$, one also has $\forall Y\in \SAT$ that $\rho,\xi(X:=Y)\vDash\Gamma$, therefore $\forall Y\in \SAT$, $\val{\ve{t}}_\rho\in\val{A}_{\xi(X:=Y)}$, then by definition of $\val{\forall X.A}_\xi$, $\val{\ve{t}}_\rho\in\val{\forall X.A}_\xi$, \ie~$\rho,\xi\vDash\ve{t}\type \forall X.A$.}
\medskip

\item\parbox{4.2cm}{
\prooftree\Gamma\Vdash\ve{t}\type A\qquad\Gamma\Vdash\ve{r}\type A
\justifies\Gamma\Vdash\ve{t}+\ve{r}\type A
\using +_I\rulesf
\endprooftree}\hspace{0.5cm}\parbox{9.2cm}{
Assume $\rho,\xi\vDash\Gamma$ in order to show $\rho,\xi\vDash\ve{t}+\ve{r}\type A$. By the induction hypothesis one has $\Gamma\vDash\ve{t}\type A$ and $\Gamma\vDash\ve{r}\type A$, so  $\val{\ve{t}}_{\rho}\in\val{A}_{\xi}$ and $\val{\ve{r}}_{\rho}\in\val{A}_{\xi}$. Since
$\val{\ve{t}+\ve{r}}_{\rho} = 
(\ve{t}+\ve{r})[\vec{x}:=\rho(\vec{x})] =
\ve{t}[\vec{x}:=\rho(\vec{x})]+\ve{r}[\vec{x}:=\rho(\vec{x})] =
\val{\ve{t}}_{\rho}+\val{\ve{r}}_{\rho}$, 
it follows from the saturation of $\val{A}_{\xi}$ that $\val{\ve{t}+\ve{r}}_{\rho}\in\val{A}_{\xi}$.}
\medskip

\item\parbox{2,2cm}{
\vspace{-0,1cm}\prooftree\Gamma\Vdash\ve{t}\type A
\justifies\Gamma\Vdash\alpha.\ve{t}\type A
\using s_I\rulesf
\endprooftree}\hspace{0.5cm}\parbox{11.2cm}{
Suppose $\rho,\xi\vDash\Gamma$ in order to show $\rho,\xi\vDash\alpha.\ve{t}\type A$. By the induction hypothesis one has $\Gamma\vDash\ve{t}\type A$, then $\val{\ve{t}}_{\rho}\in\val{A}_{\xi}$. Since
$\val{\alpha.\ve{t}}_{\rho} =
(\alpha.\ve{t})[\vec{x}:=\rho(\vec{x})] =
\alpha.\ve{t}[\vec{x}:=\rho(\vec{x})] =
\alpha.\val{\ve{t}}_{\rho}$,
it follows from the saturation of $\val{A}_{\xi}$ that $\val{\alpha.\ve{t}}_{\rho}\in\val{A}_{\xi}$.\qed}
\end{enumerate}

\subsection{Proof of Lemma~\ref{lem:corresp}}\label{proof:corresp}
We proceed by induction on the derivation of $\Gamma\vdash\ve{t}\type T$.
\begin{enumerate}
\item\parbox{2.9cm}{
\prooftree
\justifies\Gamma,{x}\type U\vdash x\type U
\using ax
\endprooftree}\hspace{0.5cm}\parbox{10.5cm}{
$(\Gamma,x\type U)\mapsf = \Gamma\mapsf,x\type U\mapsf$, so by $ax\rulesf$, $(\Gamma,x\type U)\mapsf\Vdash x\type U\mapsf$.}
\medskip

\item\parbox{2cm}{
\prooftree
\justifies\Gamma\vdash\ve{0}\type\tzero
\using ax_\tzero
\endprooftree}\hspace{0.5cm}\parbox{10.4cm}{
By ${ax_0}\rulesf$, $\Gamma\mapsf\Vdash\ve{0}\type A$ for any $A\in\Tsf$, so take $A=\tzero\mapsf$.}
\medskip

\item\parbox{6.2cm}{
\prooftree\Gamma\vdash\ve{t}\type\alpha.(U\to T)\qquad\Gamma\vdash\ve{r}\type\beta.U
\justifies\Gamma\vdash\ve{t}\ve{r}\type(\alpha\times\beta).T
\using\to_E
\endprooftree}\hspace{0.5cm}\parbox{7.2cm}{
By the induction hypothesis $\Gamma\mapsf\Vdash\ve{t}\type U\mapsf\to T\mapsf$ and $\Gamma\mapsf\Vdash\ve{r}\type U\mapsf$, so by the rule $\to_E\rulesf$, $\Gamma\mapsf\Vdash\ve{t}\ve{r}\type T\mapsf = ((\alpha\times\beta).T)\mapsf$.}
\medskip

\item\parbox{3.4cm}{
\prooftree\Gamma,x\type U\vdash\ve{t}\type T
\justifies\Gamma\vdash\lambda x\,\ve{t}\type U\to T
\using\to_I
\endprooftree}\hspace{0.5cm}\parbox{10cm}{
By the induction hypothesis $\Gamma\mapsf,x\type U\mapsf\Vdash\ve{t}\type T\mapsf$, so by the rule $\to_I\rulesf$, $\Gamma\mapsf\Vdash\lambda x\,\ve{t}\type U\mapsf\to T\mapsf = (U\to T)\mapsf$.}
\medskip

\item\parbox{3cm}{
\prooftree\Gamma\vdash\ve{t}\type\forall X.T
\justifies\Gamma\vdash\ve{t}\type T[U/X]
\using\forall_E
\endprooftree}\hspace{0.5cm}\parbox{10.4cm}{
By the induction hypothesis $\Gamma\mapsf\Vdash\ve{t}\type(\forall X.T)\mapsf=\forall X.T\mapsf$, so by the rule $\forall_E\rulesf$, $\Gamma\mapsf\Vdash\ve{t}\type T\mapsf[U\mapsf/X]$.}
\medskip

\item\parbox{2.5cm}{
\prooftree\Gamma\vdash\ve{t}\type T
\justifies\Gamma\vdash\ve{t}\type\forall X.T
\using\forall_I
\endprooftree}\hspace{0.5cm}\parbox{10.9cm}{
By the induction hypothesis $\Gamma\mapsf\Vdash\ve{t}\type T\mapsf$, so by the rule $\forall_I\rulesf$, $\Gamma\mapsf\Vdash\ve{t}\type \forall X.T\mapsf=(\forall X.T)\mapsf$. (Note that $FV(\Gamma)=FV(\Gamma\mapsf)$).}
\bigskip

\item\parbox{4.8cm}{
\prooftree\Gamma\vdash\ve{t}\type\alpha.T\qquad\Gamma\vdash\ve{r}\type\beta.T
\justifies\Gamma\vdash\ve{t}+\ve{r}\type(\alpha+\beta).T
\using +_I
\endprooftree}\hspace{0.5cm}\parbox{8.6cm}{Given in the main text of the paper.}
\medskip

\item\parbox{2.5cm}{
\prooftree\Gamma\vdash\ve{t}\type T
\justifies\Gamma\vdash\alpha.\ve{t}\type\alpha.T
\using s_I
\endprooftree}\hspace{0.5cm}\parbox{10.9cm}{
By the induction hypothesis $\Gamma\mapsf\Vdash\ve{t}\type T\mapsf$, so by the rule $s_I\rulesf$, $\Gamma\mapsf\Vdash\alpha.\ve{t}\type T\mapsf=(\alpha.T)\mapsf$.\qed}
\end{enumerate}

\subsection{Proof of Theorem~\ref{thm:confluence}(3)}\label{proof:confluence}
Before proving this theorem, we need to prove the following properties:
\begin{lem}\label{lem:replacealgrule}~
\begin{enumerate}[\em(1)]
 \item\label{it:replacebasealgrule} Let $\ve t$ and $\ve r$ be any terms. If $\ve t\to_a\ve r$, then for any base term $\ve b$, $\ve t[\ve b/x]\to_a\ve r[\ve b/x]$.
 \item\label{it:replaceargalgrule} Let $\ve b_1$ and $\ve b_2$ be base terms. If $\ve b_1\to_a\ve b_2$, then for any term $\ve t$, $\ve t[\ve b_1/x]\to_a^*\ve t[\ve b_2/x]$.
\end{enumerate}
\end{lem}

\proof\hfill
\begin{enumerate}[(1)]
\item  We proceed by structural induction on $\ve t$.
\begin{enumerate}[(a)]
 \item $\ve t$ cannot be a variable or $\ve 0$, since it reduces.
 \item $\ve t=\lambda x\,\ve t'$ and $\ve r=\lambda x\,\ve r'$, with $\ve t'\to_a \ve r'$. Then $\lambda x\,\ve t'[\ve b/x]=\lambda x\,(\ve t'[\ve b/x])$, which by the induction hypothesis $\to_a$-reduces to $\lambda x\,(\ve r[\ve b/x])=\lambda x\,\ve r[\ve b/x]$.
 \item $\ve t=\alpha.\ve t'$. We proceed now by cases:
  \begin{enumerate}[(i)]
   \item $\alpha=0$ and $\ve r=0$, then $(0.\ve t')[\ve b/x]=0.\ve t'[\ve b/x]\to_a\ve 0=\ve 0[\ve b/x]$.
   \item $\alpha=1$ and $\ve r=\ve t'$, then $(1.\ve t')[\ve b/x]=1.\ve t'[\ve b/x]\to_a\ve t'[\ve b/x]$.
   \item $\ve t=\ve 0$ and $\ve r=\ve 0$, then $(\alpha.\ve 0)[\ve b/x]=\alpha.\ve 0[\ve b/x]=\alpha.\ve 0\to_a\ve 0=\ve 0[\ve b/x]$.
   \item $\ve t'=\beta.\ve t''$ and $\ve r=(\alpha\times\beta).\ve t''$, then $(\alpha.\beta.\ve t'')[\ve b/x]=\alpha.\beta.\ve t''[\ve b/x]\to_a(\alpha\times\beta).\ve t''[\ve b/x]=((\alpha\times\beta).\ve t'')[\ve b/x]$.
   \item $\ve t'=\ve t_1+\ve t_2$ and $\ve r=\alpha.\ve t_1+\alpha.\ve t_2$, then $(\alpha.(\ve t_1+\ve t_2))[\ve b/x]=\alpha.(\ve t_1[\ve b/x]+\ve t_2[\ve b/x])\to_a\alpha.\ve t_1[\ve b/x]+\alpha.\ve t_2[\ve b/x]=(\alpha.\ve t_1+\alpha.\ve t_2)[\ve b/x]$.
   \item $\ve t'\to_a \ve r'$, and $\ve r=\alpha.\ve r'$, then $(\alpha.\ve t')[\ve b/x]=\alpha.\ve t'[\ve b/x]$, which by the induction hypothesis $\to_a$-reduces to $\alpha.\ve r'[\ve b/x]=(\alpha.\ve r')[\ve b/x]$.
  \end{enumerate}
 \item $\ve t=\ve t_1+\ve r_2$. We proceed now by cases:
  \begin{enumerate}[(i)]
   \item $\ve t_2=0$ and $\ve r=\ve t_1$, then $(\ve t_1+\ve 0)[\ve b/x]=\ve t_1[\ve b/x]+0\to_a\ve t_1[\ve b/x]$.
   \item $\ve t_1=\alpha.\ve t'$, $\ve t_2=\beta.\ve t'$ and $\ve r=(\alpha+\beta).\ve t'$, then $(\alpha.\ve t'+\beta.\ve t')[\ve b/x]=\alpha.\ve t'[\ve b/x]+\beta.\ve t'[\ve b/x]\to_a(\alpha+\beta).\ve t'[\ve b/x]=((\alpha+\beta).\ve t')[\ve b/x]$.
   \item $\ve t_1=\alpha.\ve t'$, $\ve t_2=\ve t'$ and $\ve r=(\alpha+1).\ve t'$. Analogous to the previous case.
   \item $\ve t_1=\ve t_2=\ve t'$, and $\ve r=(1+1).\ve t'$. Analogous to the previous case.
   \item $\ve t_1\to_a \ve t'$ and $\ve r=\ve t'+\ve t_2$, then $(\ve t_1+\ve t_2)[\ve b/x]=\ve t_1[\ve b/x]+\ve t_2[\ve b/x]$, which by the induction hypothesis $\to_a$-reduces to $\ve t'[\ve b/x]+\ve t_2[\ve b/x]=(\ve t'+\ve t_2)[\ve b/x]$.
   \item $\ve t_2\to_a\ve t'$ and $\ve r+\ve t_1+\ve t'$. Analogous to the previous case.
  \end{enumerate}
 \item $\ve t=\ve t_1\ve t_2$. We proceed now by cases:
  \begin{enumerate}[(i)]
   \item $\ve t_1=\ve t_{11}+\ve t_{12}$ and $\ve r=\ve t_{11}\ve
     t_2+\ve t_{12}\ve t_2$, then $((\ve t_{11}+\ve t_{12})\ve t_2)[\ve b/x]
  =(\ve t_{11}[\ve b/x]+\ve t_{12}[\ve b/x])\ve t_2[\ve b/x]\to_a\ve
  t_{11}[\ve b/x]\ve t_2[\ve b/x]+\ve t_{12}[\ve b/x]\ve t_2[\ve
    b/x]=(\ve t_{11}\ve t_2)[\ve b/x]+\\(\ve t_{12}\ve t_2)[\ve b/x]=(\ve
  t_{11}\ve t_2+\ve t_{12}\ve t_2)[\ve b/x]$.
   \item $\ve t_2=\ve t_{21}+\ve t_{22}$ and $\ve t=\ve t_1\ve t_{21}+\ve t_1\ve t_{22}$. Analogous to the previous case.
   \item $\ve t_1=\alpha.\ve t'$ and $\ve r=\alpha.\ve t'\ve t_2$, then $((\alpha.\ve t')\ve t_2)[\ve b/x]=\alpha.\ve t'[\ve b/x]\ve t_2[\ve b/x]\to_a\\\alpha.\ve t'[\ve b/x]\ve t_2[\ve b/x]=(\alpha.\ve t'\ve t_2)[\ve b/x]$.
   \item $\ve t_2=\alpha.\ve t'$ and $\ve r=\alpha.\ve t_1\ve t'$. Analogous to the previous case.
   \item $\ve t_1=\ve r=\ve 0$, then $(\ve 0\ve t_2)[\ve b/x]=\ve 0\ve t_2[\ve b/x]\to_a\ve 0=\ve 0[\ve b/x]$.
   \item $\ve t_2=\ve r=\ve 0$. Analogous to the previous case.
   \item $\ve t_1\to_a\ve t'$ and $\ve r=\ve t'\ve t_2$, then $(\ve t_1\ve t_2)[\ve b/x]=\ve t_1[\ve b/x]\ve t_2[\ve b/x]$, which by the induction hypothesis $\to_a$-reduces to $\ve t'[\ve b/x]\ve t_2[\ve b/x]=(\ve t'\ve t_2)[\ve b/x]$.
   \item $\ve t_2\to_a\ve t'$ and $\ve r=\ve t_1\ve t'$. Analogous to the previous case.
  \end{enumerate}
\end{enumerate}
\item We proceed by structural induction over $\ve t$.
  \begin{enumerate}[(i)]
	\item $\ve t=x$, then $x[\ve b_1/x]=\ve b_1\to_a\ve b_2=x[\ve b_2/x]$.
	\item $\ve t=y$, then $y[\ve b_1/x]=y=y[\ve b_2/x]$.
	\item $\ve t=\ve 0$, analogous to the previous case.
	\item $\ve t=\lambda y\,\ve t'$, then $(\lambda y\,\ve t')[\ve b_1/x]=\lambda y\,\ve t'[\ve b_1/x]$. By the induction hypothesis $\ve t'[\ve b_1/x]\to_a^*\ve t'[\ve b_2/x]$, so $\lambda y\,\ve t'[\ve b_1/x]\to_a^*\lambda y\,\ve t'[\ve b_2/x]$.
	\item $\ve t=\ve t_1\ve t_2$. Then $(\ve t_1\ve t_2)[\ve b_1/x]=\ve t_1[\ve b_1/x]\ve t_2[\ve b_1/x]$. By the induction hypothesis $\ve t_1[\ve b_1/x]\to_a^*\ve t_1[\ve b_2/x]$ and $\ve t_2[\ve b_1/x]\to_a^*\ve t_2[\ve b_2/x]$, so $\ve t_1[\ve b_1/x]\ve t_2[\ve b_1/x]\to_a^*\ve t_1[\ve b_2/x]\ve t_2[\ve b_2/x]=(\ve t_1\ve t_2)[\ve b_2/x]$.
	\item $\ve t=\ve t_1+\ve t_2$. Then $(\ve t_1+\ve t_2)[\ve b_1/x]=\ve t_1[\ve b_1/x]+\ve t_2[\ve b_1/x]$. By the induction hypothesis $\ve t_1[\ve b_1/x]\to_a^*\ve t_1[\ve b_2/x]$ and $\ve t_2[\ve b_1/x]\to_a^*\ve t_2[\ve b_2/x]$, so one has $\ve t_1[\ve b_1/x]+\ve t_2[\ve b_1/x]\to_a^*\ve t_1[\ve b_2/x]+\ve t_2[\ve b_2/x]=(\ve t_1+\ve t_2)[\ve b_2/x]$.
	\item $\ve t=\alpha.\ve t'$. Then $(\alpha.\ve t')[\ve b_1/x]=\alpha.\ve t'[\ve b_1/x]$. By the induction hypothesis one has $\ve t'[\ve b_1/x]\to_a^*\ve t'[\ve b_2/x]$, so $\alpha.\ve t'[\ve b_1/x]\to_a^*\alpha.\ve t'[\ve b_2/x]=(\alpha.\ve t')[\ve b_2/x]$.\qed
  \end{enumerate}
\end{enumerate}

\noindent Now, we can prove Theorem~\ref{thm:confluence}(3), using this property.
\proof Let $\ve t\to_a\ve{u}$ and $\ve t\to_\beta\ve r$. We proceed by structural induction on $\ve t$.
\begin{enumerate}[(1)]
  \item $\ve t$ cannot be a variable or $\ve 0$, since it reduces.
  \item $\ve t=\lambda x\,\ve t_1$. Then $\ve{u}=\lambda x\,\ve{u}_1$
    and $\ve r=\lambda x\,\ve r_1$ with $\ve t_1\to_a\ve{u}_1$ and
    $\ve t_1\to_\beta\ve r_1$. Then by the induction hypothesis there
    exists $\ve t_1'$ such that $\ve{u}_1\to_\beta\ve t_1'$ and $\ve
    r_1\to_a\ve t_1'$. Take $\ve t'=\lambda x\,\ve t_1'$.
  \item $\ve t=\ve t_1\ve t_2$. We proceed now by cases:
    \begin{enumerate}[(a)]
	\item $\ve r=\ve r_1\ve t_2$ where $\ve t_1\to_\beta\ve r_1$. We proceed now by cases:
	  \begin{enumerate}[(i)]
	    \item $\ve{u}=\ve{u}_1\ve t_2$ where $\ve t_1\to_a\ve{u}_1$. Then by the induction hypothesis, there exists $\ve t_1'$ such that $\ve r_1\to_a\ve t_1'$ and $\ve{u}_1\to_\beta\ve t_1'$. Take $\ve t'=\ve t_1'\ve t_2$.
	    \item $\ve{u}=\ve t_1\ve{u}_2$ where $\ve t_2\to_a\ve{u}_2$. Take $\ve t'=\ve r_1\ve{u}_2$.
	  \end{enumerate}
	\item $\ve r=\ve t_1\ve r_2$ where $\ve t_2\to_\beta\ve r_2$. Analogous to the previous case.
	\item $\ve r=\ve r_1[\ve t_2/x]$ where $\ve t_1=\lambda x\,\ve r_1$ and $\ve t_2$ is a base term. We proceed now by cases:
	  \begin{enumerate}[(i)]
	    \item $\ve{u}=(\lambda x\,\ve{u}_1)\ve t_2$ where $\ve r_1\to_a\ve{u}_1$. 
	    By Lemma~\ref{lem:replacealgrule}(\ref{it:replacebasealgrule}), $\ve r=\ve r_1[\ve t_2/x]\to_a\ve u_1[\ve t_2/x]$, and note that $\ve{u}\to_\beta\ve{u}_1[\ve t_2/x]$. 
	    \item $\ve{u}=(\lambda x\,\ve r_1)\ve{u}_2$ where $\ve t_2\to_a\ve{u}_2$. 
	    By Lemma~\ref{lem:replacealgrule}(\ref{it:replaceargalgrule}), $\ve r=\ve r_1[\ve t_2/x]\to_a^*\ve r_1[\ve u_2/x]$, and note that $\ve u\to_\beta\ve r_1[\ve u_2/x]$.
	  \end{enumerate}
    \end{enumerate}
    \item $\ve t=\ve t_1+\ve t_2$. We proceed now by cases:
    \begin{enumerate}[(a)]
    \item $\ve{u}=(\alpha+\beta).\ve{u}'$ with $\ve t_1=\alpha.\ve{u}'$ and $\ve t_2=\beta.\ve{u}'$. We proceed now by cases:
      \begin{enumerate}[(i)]
	\item $\ve r=\alpha.\ve r'+\beta.\ve{u}'$ with $\ve{u}'\to_\beta\ve r'$. Then note that $\ve{u}=(\alpha+\beta).\ve{u}'\to_\beta(\alpha+\beta).\ve r'$ and $\alpha.\ve r'+\beta.\ve{u}'\to_\beta\alpha.\ve r'+\beta.\ve r'\to_a(\alpha+\beta).\ve r'$.
	\item $\ve r=\alpha.\ve{u}'+\beta.\ve r'$ with $\ve{u}'\to_\beta\ve r'$. Analogous to the previous case.
      \end{enumerate}
    \item $\ve{u}=(\alpha+1).\ve{u}'$ with $\ve t_1=\alpha.\ve{u}'$ and $\ve t_2=\ve{u}'$. Analogous to the previous case.
    \item $\ve{u}=(1+1).\ve{u}'$ with $\ve t_1=\ve t_2$. Analogous to the previous case.
    \item $\ve{u}=\ve t_1$ with $\ve t_2=\ve 0$. Then the only possibility for $\ve r$ is to be $\ve r'+\ve 0$ where $\ve t_1\to_\beta\ve r'$. Then $\ve r'+\ve 0\to_a\ve r'$, which closes the case.
  \end{enumerate}
  \item $\ve t=\alpha.\ve t_1$. The only possibility for $\ve r$ is to be $\alpha.\ve r'$ where $\ve t_1\to_\beta\ve r'$. We proceed now by cases:
    \begin{enumerate}[(a)]
      \item $\ve{u}=(\alpha\times\beta).\ve{u}_1$ with $\ve t_1=\beta.\ve{u}_1$, then $\ve r'=\beta.\ve r''$ with $\ve{u}_1\to_\beta\ve r''$. Then $(\alpha\times\beta).\ve{u}_1\to_\beta(\alpha\times\beta).\ve r''$ and $\alpha.(\beta.\ve r'')\to_a(\alpha\times\beta).\ve r''$.
      \item $\ve{u}=\alpha.\ve{u}_1+\alpha.\ve{u}_2$ with $\ve t_1=\ve{u}_1+\ve{u}_2$. We proceed now by cases:
      \begin{enumerate}[(i)]
	\item $\ve r'=\ve r_1+\ve{u}_2$ with $\ve{u}_1\to_\beta\ve r_1$, then $\alpha.(\ve r_1+\ve{u}_2)\to_a\alpha.\ve r_1+\alpha.\ve{u}_2$ and $\alpha.\ve{u}_1+\alpha.\ve{u}_2\to_\beta\alpha.\ve r_1+\alpha.\ve{u}_2$.
	\item $\ve r'=\ve{u}_1+\ve r_2$ with $\ve{u}_2\to_\beta\ve r_2$, analogous to the previous case.
      \end{enumerate}
      \item $\ve{u}=\ve t_1$ with $\alpha=1$. Note that $1.\ve r'\to_a \ve r'$.
      \item $\ve{u}=\ve 0$ with $\alpha=0$. Note that $\alpha.\ve r'\to_a\ve 0$.
      \item $\ve{u}=\ve 0$ with $\ve t_1=\ve 0$. Absurd since $\ve t_1\to_\beta\ve r'$. \qed
    \end{enumerate}
\end{enumerate}

\subsection{Proof of Lemma~\ref{lem:R2}}\label{proof:R2}
Structural induction on $\ve{t}_1$.
\begin{enumerate}
  \item $\ve{t}_1=x$. Done.
  \item $\ve{t}_1=\lambda x\,\ve{r}$, then $\ve{t}_1\ve{t}_2\to \ve{r}[\ve{t}_2/x]$, which is a contradiction.
  \item $\ve{t}_1=\ve{0}$, then $\ve{t}_1\ve{t}_2\to\ve{0}$, which is a contradiction.
  \item $\ve{t}_1=\alpha.\ve{r}$, then $\ve{t}_1\ve{t}_2\to\alpha.\ve{r}\ve{t}_2$, which is a contradiction.
  \item $\ve{t}_1=\ve{r}+\ve{u}$, then $\ve{t}_1\ve{t}_2\to\ve{r}\ve{t}_2+\ve{u}\ve{t}_2$, which is a contradiction.
  \item $\ve{t}_1=\ve{u}_1\ve{u}_2$, then by the induction hypothesis $\ve{u}_1=x\vec{\ve{r}}$, so $\ve{u}_1\ve{u}_2=x\vec{\ve{r}}'$, where $\vec{\ve{r}}'=\vec{\ve{r}}, \ve{u}_2$. \qed
\end{enumerate}

\subsection{Proof of Lemma~\ref{lem:R3}}\label{proof:R3}

Fist we need some remarks.
\begin{rem}\label{rem}~
 \begin{enumerate}[(1)]
  \item Variables can only have type $T\in\Tsf$, since contexts have only classic types in the type system $\mathcal{B}$, and forall eliminations do not introduce types with scalars.
  \item\label{it:rem} If $A\in\Tsf$ and $U$ is a unit type, with $\alpha.U\equiv\beta.A$, then it is easy to check  that $U\in\Tsf$, $\alpha=\beta$ and $U\equiv A$, since there is no way using the equivalences to remove a scalar at the right of an arrow.
 \end{enumerate}
\end{rem}

\noindent Now, we prove Lemma~\ref{lem:R3}
\proof I induction on the derivation of $\Gamma\vdash x\vec{\ve{r}}\type T$. We only consider the cases ending with an application.
\begin{enumerate}[(1)]
\item\parbox{5.6cm}{
	\prooftree
	\Gamma\vdash x \type U\to T
	\qquad
	\Gamma\vdash\ve{t}\type\beta.U
	\justifies\Gamma\vdash x\ve{t}\type\beta.T
	\using\to_E
	\endprooftree}\hspace{0.5cm}\parbox{7.8cm}{
	Since $U\to T\in\Tsf$, then $T\in\Tsf$.}
	\medskip

\item\parbox{6.8cm}{
	\prooftree
	\Gamma\vdash x\vec{\ve{r}}\type\alpha.(U\to T)
	\qquad
	\Gamma\vdash\ve{t}\type\beta.U
	\justifies\Gamma\vdash (x\vec{\ve{r}})\ve{t}\type(\alpha\times\beta).T
	\using\to_E
	\endprooftree}\hspace{0.5cm}\parbox{6.6cm}{
	Then by the induction hypothesis, $U\to T\in\Tsf$, so $T\in\Tsf$.}
	\medskip
	
\item\parbox{3.7cm}{
	\prooftree\Gamma\vdash x\vec{\ve{r}}\type \forall X.T
	\justifies\Gamma\vdash x\vec{\ve{r}}\type T[A/X]
	\using\forall_E
	\endprooftree}\hspace{0.5cm}\parbox{9.7cm}{
	Then by the induction hypothesis there exist $B\in\Tsf$ and $\alpha\in\Sc$ such that $\forall X.T\equiv\alpha.B$, so there exists $C\in\Tsf$ such that $T\equiv\alpha.C$, then $T[A/X]\equiv(\alpha.C)[A/X]\equiv\alpha.C[A/X]$. Note that $C[A/X]\in\Tsf$.}
  \medskip
  
\item\parbox{3.2cm}{
	\prooftree\Gamma\vdash x\vec{\ve{r}}\type T
	\justifies\Gamma\vdash x\vec{\ve{r}}\type \forall X.T
	\using\forall_I
	\endprooftree}\hspace{0.5cm}\parbox{10.2cm}{
	Then by the induction hypothesis there exist $A\in\Tsf$ and $\alpha\in\Sc$ such that $T\equiv\alpha.A$, so $\forall X.T\equiv\forall X.\alpha.A\equiv\alpha.\forall X.A$. \qed}
\end{enumerate}

\subsection{Proof of Lemma~\ref{lem:scalarsnotaddedbysubsumption}}\label{proof:scalarsnotaddedbysubsumption}
First note that if $A\in\Tsf$ and $\alpha.T\equiv\alpha.A$, then $T=A$ or $T=1.A$: The only equivalence rule involving a type that could be in $\Tsf$ is $1.T\equiv T$, so if $\alpha.T\equiv\alpha.A$, $T$ must be either $1.A$ or $A$.

Now we prove the Lemma for $\succ_{X,\Gamma}^{\ve t}$, and in the light of Lemma~\ref{lem:scalarskeeporder}, we can remove the scalar $\alpha$. We proceed now by cases:
\begin{iteMize}{$\bullet$}
 \item $A\equiv\forall X.T$, then either $\forall X.T=A$, so $T\in\Tsf$, or $\forall X.T=1.A$, in which case $T=1.B$ with $B\in\Tsf$.
 \item $T\equiv\forall X.R$ and $A\equiv R[U/X]$. Since $R[U/X]\in\Tsf$, then $R\in\Tsf$ and so $\forall X.R\in\Tsf$, then either $T\in\Tsf$ or $T\equiv 1.B$ with $B\in\Tsf$.\qed
\end{iteMize}

\subsection{Proof of Theorem~\ref{thm:weight1}}\label{proof:weight1}
Instead, we prove the most general case: If $\Gamma\vdash\ve{t}\type\alpha.A$ then $\omega(\nf{\ve{t}})=\alpha$, by structural induction on $\nf{\ve{t}}$. We take $\Gamma\vdash\nf{\ve{t}}\type\alpha.A$, which is true by Theorem~\ref{thm:subjectreduction}.

\begin{enumerate}[(1)]
  \item $\nf{\ve{t}}=\ve{0}$. Then $\omega(\nf{\ve{t}})=0$. In addition, by Lemma~\ref{lem:type0}, $\alpha.A\equiv\tzero$, so by Theorem~\ref{thm:uniqscalar}, $\alpha=0$.
  \item $\nf{\ve{t}}=x$ or $\nf{\ve{t}}=\lambda x\,\ve{t}'$. Then $\omega(\nf{\ve{t}})=1$. In addition, by Lemma~\ref{lem:baseterms}, $\alpha=1$.
  \item $\nf{\ve{t}}=\gamma.\ve{t}'$. 
Given in the main text of the paper.
  \item $\nf{\ve{t}}=\ve{t}_1+\ve{t}_2$. Then $\omega(\nf{\ve{t}})=\omega(\ve{t}_1)+\omega(\ve{t}_2)$. By Lemma~\ref{lem:gen-sum}, there exist $\sigma, \phi\in\Sc$ such that $\Gamma\vdash\ve{t}_1\type\sigma.A$ and $\Gamma\vdash\ve{t}_2\type\phi.A$ with $\sigma+\phi=\alpha$.
  Then by the induction hypothesis $\omega(\ve{t}_1)=\sigma$ and $\omega(\ve{t}_2)=\phi$, so $\omega(\ve{t}_1)+\omega(\ve{t}_2)=\sigma+\phi=\alpha$.
  \item $\nf{\ve{t}}=\ve{t}_1\ve{t}_2$. 
Given in the main text of the paper.
\qed
\end{enumerate}

\end{document}